\documentclass[useAMS,usenatbib]{mn2e}
\usepackage{multirow}
\usepackage{multicol}
\usepackage{graphicx}
\usepackage{color}
\title[Outer density profiles of Galactic globular clusters]{Outer density profiles of 19 Galactic globular clusters from deep and wide-field imaging}
\author[Carballo-Bello et al.]
{Julio A. Carballo-Bello$^{1,2}$
\thanks{E-mail: jacb@iac.es},
Mark Gieles$^{3}$, Antonio Sollima$^{4}$, Sergey Koposov$^{3}$,
\newauthor
  David Mart\'inez-Delgado$^{5}$ and Jorge Pe\~narrubia$^{3}$
\\
\\
$^{1}$Instituto de Astrof\'isica de Canarias (IAC), V\'ia L\'actea s/n, La Laguna E-38205, S/C de Tenerife, Spain\\
$^{2}$Departamento de Astrof\'isica, Universidad de La Laguna, La Laguna E-38205, S/C de Tenerife, Spain\\
$^{3}$Institute of Astronomy, University of Cambridge,  CB3 0HA Cambridge, United Kingdom\\
$^{4}$INAF- Observatorio Astronomico di Padova, I35122 Padova, Italy\\
$^{5}$Max Planck Institut f\"ur Astronomie, D69117 Heidelberg, Germany \\
}

\newcommand{\gyr}{{\rm Gyr}}
\newcommand{\myr}{{\rm Myr}}
\newcommand{\mmean}{\bar{m}}
\newcommand{\rhoh}{\rho_{\rm h}}
\newcommand{\rc}{r_{\rm c}}
\newcommand{\reff}{r_{\rm eff}}
\newcommand{\rh}{r_{\rm h}}
\newcommand{\rt}{r_{\rm edge}}
\newcommand{\rg}{R_{\rm G}}
\newcommand{\sm}{M_{\odot}}
\newcommand{\trh}{t_{\rm rh}}

\begin{document}

\pagerange{\pageref{firstpage}--\pageref{lastpage}} \pubyear{2011}

\maketitle

\label{firstpage}

\begin{abstract}
Using deep photometric data from WFC@INT and WFI@ESO2.2m we measure the outer 
number density profiles of 19 stellar clusters located in the inner region of the 
Milky Way halo (within a Galactocentric distance range of 10--30 kpc) in order to 
assess the impact of internal and external dynamical processes on the spatial 
distribution of stars. Adopting power-law fitting templates, with index $-\gamma$ in the outer region, we find that 
the clusters in our sample can be divided in two groups: a group of massive clusters ($ \ge  10^5\,\sm$) that has 
relatively flat profiles with $2.5 < \gamma < 4$ and a group of low-mass clusters ($ \le  10^5\,\sm$), with steep profiles ($\gamma > 4$) and clear signatures of interaction with the Galactic tidal field. We refer to these two groups as 'tidally unaffected' and 'tidally affected', respectively.
Our results also show a clear trend between the 
slope of the outer parts and the half-mass density of these systems, which 
suggests that the outer density profiles may retain key information on the 
dominant processes driving the dynamical evolution of Globular Clusters. 
\end{abstract}


\section{Introduction}

Globular clusters (GCs) have been found in nearly all galaxies and they have 
been considered the fossil records of the  formation of their 
host galaxy \citep[e.g.][]{1993ASPC...48...38Z,1993MNRAS.264..611Z,2006ARA&A..44..193B,2009ApJ...694.1498M}.
In the Milky Way roughly 150 GCs are currently
known \citep{1996AJ....112.1487H, 2010arXiv1012.3224H} and they have a
long history of being used in the context of stellar
evolution \citep[e.g.][]{1998A&A...337..403B}, the structure and
chemical evolution of the Milky
Way \citep[e.g.][respectively]{1983AJ.....88..338I,1978ApJ...225..357S}
and the dynamical evolution of collisional
systems \citep{1992PASP..104..981H,1997A&ARv...8....1M}.

The radial number density profiles of GCs can be used to
learn about their evolution because the structure and evolution of
clusters are closely linked \citep{1961AnAp...24..369H}. In recent
years, much attention has gone to the inner parts of the surface
brightness profile to quantify the number of clusters that has
undergone core collapse \citep{1986ApJ...305L..61D} and to look for
the presence of massive black
holes \citep{2006AJ....132..447N}. Especially for the densest clusters,
these studies require the high angular resolution of space based observatories, such as
the Hubble Space Telescope, to resolve the dense core
into individual stars. On the other hand, the spatial distribution of the
bulk of the stars is determined by the interplay between internal and
external dynamical processes. Internal relaxation processes tend to
populate the outer regions of the cluster with low-mass stars, whereas
the external tidal field of the host galaxy will strip some of the
stars at large radii. Hence, studies of the entire profile are useful
to gain insight on the overall evolution of individual
clusters. Since such studies require a wide field of view, they are
usually done from the ground \citep[][hereafter TGK95]{1995AJ....109..218T}.

Several self-consistent (static) models for clusters exist, using
different assumptions for the distribution
function \citep[e.g.][]{1911MNRAS..71..460P,1956MNRAS.116..288W,1963MNRAS.126..499M,1966AJ.....71...64K,1975AJ.....80..175W}. 
The surface brightness profile is often used to say something about the
correctness of a model, but unfortunately it does not always provide
enough information to constrain the underlying model, i.e. without the
use of additional information such as stellar velocities. This was
already noted by \citet{1961AnAp...24..369H} who realised that different models 
could successfully describe the surface brightness profile
 of a GC despite their different underlying physical assumptions.

Most of these models are based on simplifying assumptions, such as
spherical symmetry, an isotropic velocity distribution, stars of the
same mass and a truncation of the distribution function.  It is,
therefore, that empirical templates are often used, i.e. simple
fitting formulas that do not (necessarily) follow from a specific
distribution function, to fit and compare surface brightness profiles
of clusters. A well-known example is the \citet{1962AJ.....67..471K}
template that is typically used for old GCs and that is
characterised by a steep drop, or truncation, of the density at the
outer edge. This radius is called the tidal radius in King's paper,
but because it does not necessarily
coincide with the
radius of the zero-velocity surface, or Jacobi radius 
\citep{2010MNRAS.401.1832B, 2010MNRAS.407.2241K}, we will refer
to the radius where King models drop to zero density as the edge
radius ($r_{\rm edge})$.
\citet[][hereafter EFF87]{1987ApJ...323...54E}  
used a continuous template (i.e. without a truncation) for relatively young
clusters in the Large Magellanic Clouds (LMC). The EFF87 template is
characterised by a flat core and a power-law envelope. These are part
of the more general family of power-law
models \citep{1996MNRAS.278..488Z}.

The cluster system of the LMC is a good example of how observations of
the structure of clusters can be used as a proxy for
evolution. Because the ages of clusters in the LMC span a range from a
few Myrs to $\sim12\,\gyr$, variations of structural parameters with
age can be studied, and when interpreting this as evolution, much can
be learned. For example, \citet{1989ApJ...347L..69E}
and \citet{2003MNRAS.338...85M} showed that the core sizes of LMC
clusters increase (on average) with age, which was also found for
clusters in the Small Magellanic
Clouds \citep[SMC;][]{2003MNRAS.338..120M} and for cluster systems
outside the Local
Group \citep{2001ApJ...563..151M,2008MNRAS.389..223B}.  Numerical
studies suggest that this expansion is largely driven by internal
evolution \citep{2008MNRAS.386...65M}, rather than by external effects
such as the details of the orbit around the centre of the
galaxy \citep{2003MNRAS.343.1025W}.

The young SMC/LMC clusters and young extra-Galactic
clusters \citep[e.g.][]{2004A&A...416..537L} are well described by the
continuous power-law models, whereas for old GCs (tidally) truncated King models are often used.  
It seems, therefore, logic to conclude that clusters form with extended haloes
and that a tidal truncation develops over time through interactions of
the cluster with the tidal field of its host galaxy.  Because the
tidal field is strongest at pericentre, it is often assumed that this
is where the truncation in the cluster density is
determined \citep{1962AJ.....67..471K,1983AJ.....88..338I}.  Numerical
works, however, indicate that the reality is more complicated.
Collision-less $N$-body simulations of stellar clusters that start
with a King-type density distribution and that move on eccentric
orbits in the Galactic potential develop an halo of ``extra-tidal''
stars as they shed a fraction of their stellar mass to
tides \citep{1999MNRAS.302..771J}. Once the unbound material escapes
and equilibrium is re-established, the outer profile of the cluster
does not exhibit the original truncation and instead approaches a
power-law \citep{1995ApJ...442..142O, 2009ApJ...698..222P}. The
surface density in that halo is shallower than for the bulk of the
stars and the location where the break occurs is an indication of the
time elapsed from the last pericentre passage \citep{2009ApJ...698..222P}.
Observational examples of this phenomenon have been discovered in Pal\,5
 \citep{2003AJ....126.2385O,2004AJ....127.2753D}, NGC\,5466 
 \citep{2006ApJ...637L..29B} and Pal\,14 \citep{2011ApJ...726...47S}.

Tidal stripping is not the only process that can generate a break in
the surface density profile. Models of star clusters that lose stars
through two-body relaxation and move on a circular orbit, i.e. without
the time-dependent tides that give rise to tidal stripping, also
develop a halo of energetically unbound stars.  These stars were
dubbed ``potential escapers" and they are trapped within the Jacobi surface
because the criterion for escape is not only based on energy
but also on angular
momentum \citep{2000MNRAS.318..753F,2001MNRAS.325.1323B}. These stars
give rise to a very similar break in the outer parts of the surface
density profile as in the tidal
stripping scenario \citep[][]{2010MNRAS.407.2241K}. Such a break was found in 
the surface density/brightness profiles of the GCs
M\,92 \citep{2000A&A...356..127T},
Pal\,13 \citep{2002ApJ...574..783C},
Whiting~1 \citep{2007A&A...466..181C} and
AM~4 \citep{2009AJ....137.3809C}. Often these breaks go together
with direct detections of the characteristic ``S-shaped" tidal
features and/or tidal
tails \citep{2006ApJ...637L..29B,2010A&A...522A..71J,2010MNRAS.408L..66N}. 
When the details of the orbit of the cluster are not available, it is
difficult, if not impossible, to disentangle the relative
contributions of tidal stripping and two-body relaxation to the formation of the break in
the surface brightness profile.

\citet[][hereafter MvM05]{2005ApJS..161..304M} 
fit different models to the surface brightness profiles of Milky Way
GCs using the data of TGK95 and
they find that the more extended \citet{1975AJ.....80..175W} and EFF87
models provide equally good and sometimes even better descriptions
than the traditional \citet{1966AJ.....71...64K} models. In
particular, they show that goodness of fit parameters indicate a
preference for more extended models whenever more data in the outer
parts are available. They refrain from a detailed interpretation and
conclude that age is not the only parameter that determines whether a
King-type (i.e. truncated) model provides a good description of the
surface brightness profile, or not.
In this study we search for the additional ingredient driving
the shape of the clusters density profiles. 

This paper is organized as follows. In \S~2 we present the sample
of Galactic GCs included in this study and a description
of the observations needed for our purposes. In \S~3, we describe
how we obtained
 the number density profiles on which this study is based
and the fitting technique adopted to derive their structural
parameters. In \S~4 the number density profiles are presented and are
interpreted in \S~5 in connection with possible external and internal
processes. In \S~6 we draw our conclusions.

\section{The sample}

The clusters included in this project are part of a study that
focused on the search for tidal streams around Galactic GCs \citep{2004ASPC..327..255M}.
For this purpose, we have observed a sample of 19 GCs in the Galactic halo between 10 and
30\,kpc in Galactocentric distance, sampling $\sim$\,56$\%$ of the total number of Galactic GCs in this distance range. 
Since photometry at low Galactic latitudes is severely hampered by the presence of field stars, we have
excluded all those objects located within 20 degrees from the Galactic plane
with the exception of NGC\,2298 and Rup\,106, which have been observed in
spite of their low Galactic latitude because of their
possible association with the Canis Mayor
stream \citep{2004ASPC..327..220B,2010MNRAS.404.1203F}. Because all the GCs in our sample lie in a 20\,kpc wide distance range, we can
study the effect of a similar external Galactic tidal
field on the structure of clusters with 
different properties.

\begin{figure}
\begin{centering}
\includegraphics[scale=0.4]{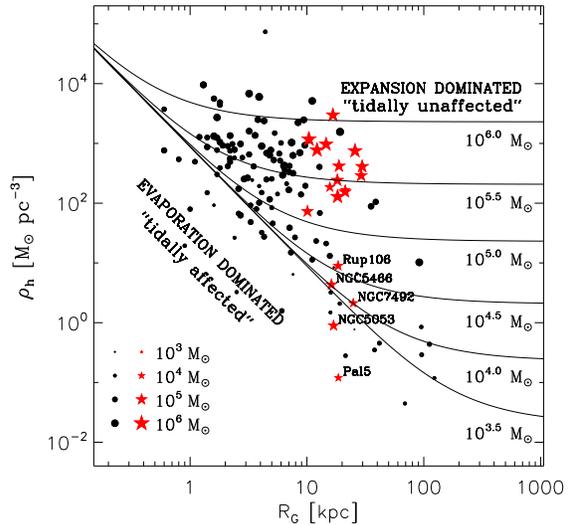} 
\caption{\small{
Half-mass densities against Galactocentric distances of the 141 globular
clusters (black dots) of
the Harris catalogue (Harris 1996). The clusters used in this study are 
highlighted with a (red) star. All
symbol sizes scale with the mass of the cluster. Cluster isochrone
relations from Gieles et al. (2011) are  over-plotted for clusters with different
masses. Clusters that align with the horizontal lines are in the tidally unaffected regime, whereas clusters that align the diagonal line are in the tidally affected regime. The exact separation between these regimes is given by equation~(\ref{eq:mass}).
}}
\label{fig:isochrones}
\end{centering}
\end{figure}

To place the sample in the context of the Milky Way GC population, we show their
half-mass densities, $\rhoh$, against the Galactocentric distance, $\rg$,  together with the values
for the rest of the GCs (Figure~\ref{fig:isochrones}). A constant mass-to-light ratio of 2 was 
used for this figure and the half-mass density is defined as $\rhoh\equiv8M/(3\pi\rh^3)$. Here $M$ is 
the mass and $\rh$ is the three-dimensional half-mass radius which is estimated from the half-light radius in 
projection, or effective radius ($r_{\rm eff}$) by correcting for the effect of projection, i.e. $\rh=(4/3)\reff$ \citep{1987degc.book.....S}.
This is a powerful diagram to study the
relation between internal and external effects who both affect cluster
properties \citep{1983AJ.....88..338I,G11}. The lines are 'cluster
isochrone relations' at an age of a Hubble time for clusters with different
masses  that evolve in a steady tidal field \citep[Appendix~B in ][]{G11}.
It assumes that two-body relaxation 
is the dominant process, which
is a valid assumption for most clusters outside
$\sim10\,$kpc \citep{1997ApJ...474..223G,1997MNRAS.289..898V}.
\citet{G11}  showed that clusters that form deeply embedded within their Roche-lobe spend roughly the first half of their lives in an
`expansion-dominated' phase and during the second half of their lives they
are in an `evaporation-dominated' phase. These authors considered relaxation driven evaporation of clusters on circular orbits. Because the evolution of clusters on eccentric orbits is affected by an additional tidal effect, i.e. tidal stripping, we introduce a more general reference to these two regimes, namely `tidally unaffected' and `tidally affected', respectively.
According to \cite{G11},  clusters that satisfy

\begin{equation}
\label{eq:mass}
M < 10^5 \sm \frac{4\,{\rm kpc}}{R_{\rm G}} 
\end{equation}
are in the tidally affected regime. This constraint is satisfied for $\sim 25\,\%$ of the clusters in our sample:
Rup\,106, NGC\,5053, NGC\,5466, Pal\,5 and NGC\,7492. 
The remaining clusters of our sample will be referred to as tidally unaffected clusters.

To derive the mass of the clusters (Table 2), we have taken the 
$V$-band mass-to-light ratios found by MvM05 and the luminosities derived from the integrated $V$ magnitudes of our profiles\footnote{To convert $V$ magnitudes into luminosities we adopted the distance moduli and reddening from \cite{2010arXiv1012.3224H} and 
$M_{V,\odot}=4.83$ \citep{1998gaas.book.....B}}. 
Moreover, trying to complete our analysis of the dynamical state of each cluster, we have used the expression

\begin{equation}
\label{eq:trh}
\trh =\frac{0.138}{\bar{m}\ln(0.11 M/\bar{m})}\bigg(\frac{ M \rh^3}{G}\bigg)^{1/2}  \rm Myr
\end{equation}

by \cite{1971ApJ...164..399S} to obtain the half-mass relaxation time
in $\myr$. In this formula, $\mmean$ is the mean stellar mass that has
been set to 0.5$\,\sm$. The constant of 0.11 was found by
\citet{1994MNRAS.268..257G} for clusters of equal mass stars.

The six-dimensional phase space information of a GC is essential to derive its orbit in the Galaxy
and evaluate its interaction with the Milky Way tidal field.
While accurate positions on the sky, distances and radial velocities are available for all the cluster of our
sample, proper motions are  difficult to obtain for these distant
objects. We only have found orbital parameters for 10 of our GCs in the compilation by
\cite{1999AJ....117.1792D}. Absolute ages included in Table\,1 are
extracted from the database by \cite{2010MNRAS.404.1203F} except for NGC\,6229 \citep{Borissova1999}. 
Coordinates and distances are taken from the Harris catalogue \citep{1996AJ....112.1487H,2010arXiv1012.3224H}.

\subsection{Observations}

\begin{table*}
\small
\begin{centering}
\begin{tabular}{rrrcccccl}
\hline
Cluster & \emph{l}($^{o}$) & \emph{b}($^{o}$) & R$_{\rm G}$(kpc) & $\reff (')$ & Age (Gyr) & R$_{\rm apo}$(kpc) & R$_{\rm peri}$(kpc) & Telescope/Run \\
\hline
\hline
\\ 
NGC\,1261  & 270.54 & -52.12 &  18.2  & 0.68& 10.24 & -- & -- & ESO2.2/Nov09\\
NGC\,1851  & 244.51 & -35.03 &  16.7  & 0.52& 9.98  & 30.4 $\pm$ 3.9 & 5.7 $\pm$ 1.1 & ESO2.2/Jan00\\
NGC\,1904  & 227.23 & -29.35 &  18.8  & 0.66& 11.14 & 19.9 $\pm$ 1.0 & 4.2 $\pm$ 1.3 & ESO2.2/Jan00\\
NGC\,2298  & 245.63 & -16.00 &  15.7  & 0.76& 12.67 & 15.3 $\pm$ 1.0 & 1.9 $\pm$ 1.4 & ESO2.2/Feb09\\
NGC\,4147  & 252.85 &  77.19 &  21.3  & 0.49& 11.39 & 25.3 $\pm$ 2.6 & 4.1 $\pm$ 2.2 & INT/May04\\
Rup\,106   & 300.88 &  11.67 &  18.5  & 1.05& 10.20 & -- & -- & ESO2.2/Feb09 \\
NGC\,4590  & 299.63 &  36.05 &  10.1  & 1.51& 11.52 & 24.4 $\pm$ 3.1 & 8.6 $\pm$ 0.3 & ESO2.2/Feb10\\
NGC\,5024  & 332.96 &  79.76 &  18.3  & 1.32& 12.67 & 36.0 $\pm$ 16.8 & 15.5 $\pm$ 1.9 & INT/May04\\
NGC\,5053  & 335.70 &  78.95 &  16.9  & 2.58& 12.29 & -- & -- & INT/May08\\
NGC\,5272  &  42.22 &  78.71 &  12.2  & 2.37& 11.39 & 13.4 $\pm$ 0.8 & 5.5 $\pm$ 0.8 & INT/May10\\
NGC\,5466  &  42.15 &  73.59 &  16.2  & 2.27& 13.57 & 57.1 $\pm$ 24.6 & 6.6 $\pm$ 1.5 & INT/May08\\ 
NGC\,5634  & 342.21 &  49.26 &  21.2  & 0.88& 11.84 & -- & -- & ESO2.2/Feb10\\
NGC\,5694  & 331.06 &  30.36 &  29.1  & 0.41& 13.44 & -- & -- & ESO2.2/Feb10\\
NGC\,5824  & 332.56 &  22.07 &  25.8  & 0.46& 12.80 & -- & -- & ESO2.2/Feb10\\
Pal\,5     &   0.85 &  45.86 &  18.6  & 2.68& 9.80  & 15.9 $\pm$ 2.5 & 2.3 $\pm$ 2.3 & INT/Jun01\\
NGC\,6229  &  73.64 &  40.31 &  29.7  & 0.36& 11.8    & -- & -- & INT/Aug09\\
NGC\,6864  &  20.30 & -25.75 &  14.6  & 0.47& 9.98  & -- & -- & ESO2.2/May10\\
NGC\,7078  &  65.01 & -27.31 &  10.4  & 1.00& 12.93 & 10.3 $\pm$ 0.7 & 5.4 $\pm$ 1.1 & INT/Jun10\\
NGC\,7492  &  53.39 & -63.48 &  24.9  & 1.13& 12.00 & -- & -- & ESO2.2/Nov09 \\
\\
\hline
\hline
\end{tabular}
\caption{\small{Sample of Galactic GCs included in this project. Coordinates 
and distances were obtained from the updated version of Harris catalogue 
(Harris 1996, 2010) while radii have been inferred from the structural 
parameters published in MVM05, except for Rup\,106, taken from Harris. 
Absolute ages are those computed by  Forbes \& Bridges (2010) except for 
NGC\,6229 (Borissova et al. 1999) and Galactocentric distances during apo- and 
pericentre of the clusters are obtained from Dinescu et al. (1999). }}
\end{centering}
\end{table*}

\nocite{1996AJ....112.1487H}
\nocite{2010arXiv1012.3224H}
\nocite{Borissova1999}

As already reported in the previous Section, the observational data used in this paper come 
from a photometric campaign aimed at searching for tidal streams around Galactic GCs. Due to
the low surface brightness of known Galactic tidal streams ($\sim$32 \,mag/arcmin$^{2}$), 
a suitable combination of wide field and
exposure time was required. We have used the Wide Field Camera (WFC) located
at the Isaac Newton Telescope at the Roque de los Muchachos
Observatory (La Palma, Spain) and the Wide Field Imager (WFI) at the
ESO\,2.2\,m telescope at La Silla Observatory (Chile). The WFC covers
34$\times$34\,arcmin with 4 identical chips while the WFI has a similar
field of view (FOV) of 34 $\times$ 33\,arcmin distributed in 8 chips.

For those clusters with angular sizes comparable to the FOVs of the
instruments, it was necessary to include an additional field to cover
the outer regions, where we focus on. The total exposure
times in $B$ and $R$ bands were 4$\times$900\,s and 6$\times$600\,s
respectively reaching a limiting magnitude of $\sim$ 5 mag below the 
turn-off in the colour-magnitude diagram (CMD). With such deep photometry we
have been able to detect cluster members in the outer parts of
these systems where the stellar density is critically low.

The images were processed using the \emph{imred} package and standard
routines in IRAF. Photometry was obtained with the PSF-fitting algorithm of 
DAOPHOT II/ALLSTAR \citep{1987PASP...99..191S}. One of the
main advantages of using this software is that it provides criteria to
reject extended objects as background galaxies so we only included
stellar-shaped objects in our final photometric catalogue. This
decision is taken based on the sharpness parameter for which we set
the limitation $|$sh$|$$\le$ 0.5. During each observing run a set of 
photometric standard stars from the catalogue by \cite{1992AJ....104..340L} 
have been observed. They have been used to derive the transformation between 
the instrumental magnitude and the standard Johnson system. We adopted the 
atmospheric extinction coefficients provided by both observatories.

\section{Number density profiles. Fitting} 

To construct the radial density profiles, we selected the bona-fide cluster 
members along the main-sequence (MS). First, we fitted the CMD of each cluster with a set of isochrones
from \cite{2008A&A...482..883M} with suitable age and metallicity from
\cite{2010MNRAS.404.1203F} and adopting the distance and reddening by \cite{2010arXiv1012.3224H}. 
We defined a box in the CMD containing all the stars with $|$($B$-$R$)$_{\rm iso}$-($B$-$R$)$_{\rm cmd}|\le$ 0.15 
and $B_{\rm to} \le  B \le  B_{\rm max}$, where $(B-R)_{\rm iso}$ and $(B-R)_{\rm cmd}$ are the isochrone and stars colours, respectively, 
 $B_{\rm to}$ the turn-off magnitude and $B_{\rm max}$ a limiting magnitude defined to be 2.23 $\leq B_{\rm max} - B_{\rm to} \leq$ 4.48 depending on the cluster. The adopted magnitude range represents a good compromise to maximize the number
of cluster objects ensuring a good level of completeness ($\phi\ge 90\%$) estimated from artificial star experiments in the most crowded region used in the analysis for a subsample of relatively dense GCs \citep[see ][]{2002AJ....123.1509B}. The comparison with the theoretical isochrones indicates that in all the GCs of our
sample, the stars that satisfy the adopted selection criterion lie in a similar
mass range, between 0.77 $\le \rm M_{\rm max} \le$ 0.86\,$\sm$ and 0.52 $\le \rm M_{\rm min} \le$ 0.69\,$\sm$, 
where M$_{\rm max}$ and M$_{\rm min}$ are the maximum and minimum stellar masses found in the 
MS of these clusters. Although the adopted selection criteria in both sharpness and location in the CMD minimize the contamination from back/foreground field stars and galaxies, the presence of some intruders in the sample is unavoidable. The fraction of contaminating objects has been estimated for a subsample of clusters for which suitable control fields located at the same Galactic latitude but $\sim$\,3\,$deg$ away from the cluster position. We estimated a fraction of outliers satisfiying the selection criteria described above of 0.1-0.3$\%$. Morever, such objects are expected to be homogeneously distributed across the field of view and should therefore not alter significantly the profile shape. So, we can can safely consider negligible the effect of such intruders in the derived profiles.

We counted all those stars included in our box that are contained in concentric 
annuli, centred in the cluster centre coordinates with fixed width in logarithmic 
scale. The number of counts per unit area has been obtained by dividing the 
number of stars that satisfy the above criteria by the corresponding area 
covered by the annuli. The area of each annulus has been corrected for the 
region not covered by our observations (gaps between the chips, borders 
truncation, ect). The error on the density has been estimated using the 
standard error propagation formula and assuming a Poisson statistic for star 
counts. 

\begin{figure}
\begin{centering}
\includegraphics[scale=0.8]{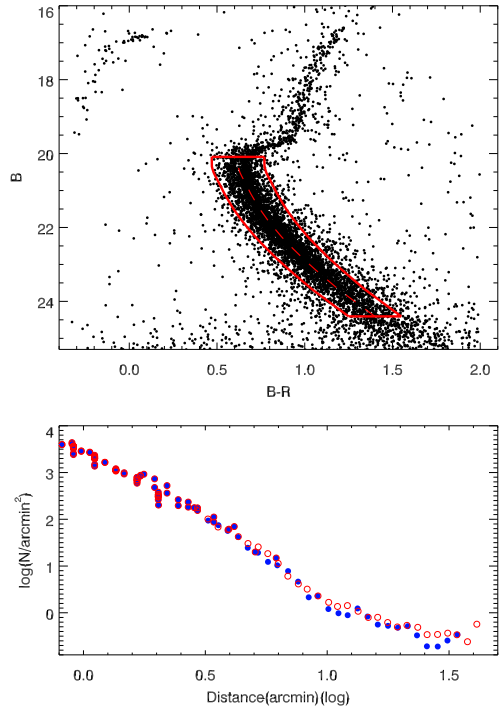}
\caption{\small{Example of MS stars selection to calculate the number density 
profile of the cluster NGC\,1904. The lower panel shows the profiles for 
different masses ranges (0.68 $\leq$ M $\leq$ 0.83\,$\sm$, open red circles; 
0.54 $\leq$ M $\leq$ 0.68\,$\sm$, solid blue circles).}}
\label{fig:sel}
\end{centering}
\end{figure}

For most of the clusters of our sample the half-mass relaxation time is smaller than
their age. It is therefore important to investigate 
the effect that mass segregation could have on the resulting profiles by splitting the MS 
in two subsamples of equal magnitude extent covering different magnitude ranges. 
We found no signs of significant differences in the obtained density profiles
for all the clusters of our sample along the entire cluster extent, with $\Delta n(r) \le 1.5\%$ (see Figure
\ref{fig:sel} for the illustrative case of NGC\,1904). The Kolmorov-Smirnov test indicates a probability of $\sim$ 98$\%$ that the radial profiles obtained for the two different subsamples of MS stars are extracted from the same population. So we can neglect the effects that mass segregation could have on our derived profiles. 

The long exposure times needed to reach a lower limiting magnitude while allowing 
to efficiently study the outer region of these GCs, causes severe 
incompleteness in the inner arcminutes which prevent us from studying the 
structure in the innermost portion of the cluster. As a solution, we have used 
the catalogue of surface brightness profiles of GCs in the Milky Way published 
by TGK95 (with the exception of Rup\,106, not included in that catalogue). For each cluster we 
have converted the TGK95 profiles into star counts adopting the relation
\begin{equation}
  n(r)=C-0.4~\mu_{V}(r)
\end{equation}
Then we fitted both our and the TGK95 dataset simultaneously,
leaving the vertical scaling factor between the two profiles, $C$, as a free parameter in the fitting model.

In the fitting procedure we optimized the combined chi-square ($\chi^{2}$)
of both datasets.
When fitting, we noticed that certain profiles in the TGK95 dataset have a 
fraction of points with very small (probably unrealistic) error-bars, which 
would completely drive the fit. Furthermore, in several cases, the data from 
TGK95 in the very inner parts of the clusters show some small scale structure 
in the profiles, which can be either intrinsic or due to systematic effects. 
In order for the fit to be determined mainly by the overall profile shape 
rather than a few points/structures in the TGK95 data, we decided to limit the 
smallest error-bar of the TGK95 dataset to 0.1 dex in density.

 The fitting model for the profiles was:
\begin{equation}
    n(r) = n_{\rm BG} + n_{\rm cl}(r) 
\end{equation}

where $n_{\rm BG}$ is the density of background/foreground sources and
$n_{\rm cl}(r)$ is the model for the cluster density profile. In this paper
we considered two different models. As in EFF87, a power-law model with parameter
$\gamma$ controlling the shape of the profile :

\begin{equation}
    n_{\rm cl}(r) = n_0 \frac{1}{(1+(r/r_0)^{2})^{\gamma/2}}
\end{equation}

and the King (1996) profile, parametrized by the $W_0$ parameter:

\begin{equation}
   n_{\rm cl}(r) = n_0\times F(r,r_0,W_0)
\end{equation}

Because \citet{1966AJ.....71...64K} models do not have an
  analytical solution for the surface density, a grid of models was
  pre-computed by numerical solving the Poisson equation.

In order to fit the density profiles, instead of standard gradient
descent methods (e.g.  Levenberg-Marquardt), we decided to employ a
Markov Chain Monte-Carlo (MCMC) technique to better explore the
possible covariance between parameters. For each cluster, we ran
multiple MCMC chains, and then from all the chains we picked up the
values of the parameters giving a best fit to the data, while the 1D
posterior distributions of the parameters gave us estimates of the
error-bars, which were often large due to the covariance between
them. Since our data combined with the TGK95 data covers very large
range of radii, for a few clusters we do see systematic deviations of
the observed profiles from our simple models. In these cases obviously
the error-bars on the parameters are only indicative. In a few cases the background 
is not well sampled by our density profiles. In our fitting procedure, the sky level is a free parameter with a flat prior between 0 and the average of a few outer points of the density profile. This parameter is sampled by the
MCMC chain and the values of all the free parameters have been determined from marginalized 1D
posteriors, so the errors on the structural parameters are in some cases significantly influenced by the
errors of the sky determination.

\section{Results}
\begin{figure*}
\begin{centering}
\includegraphics[scale=1.3]{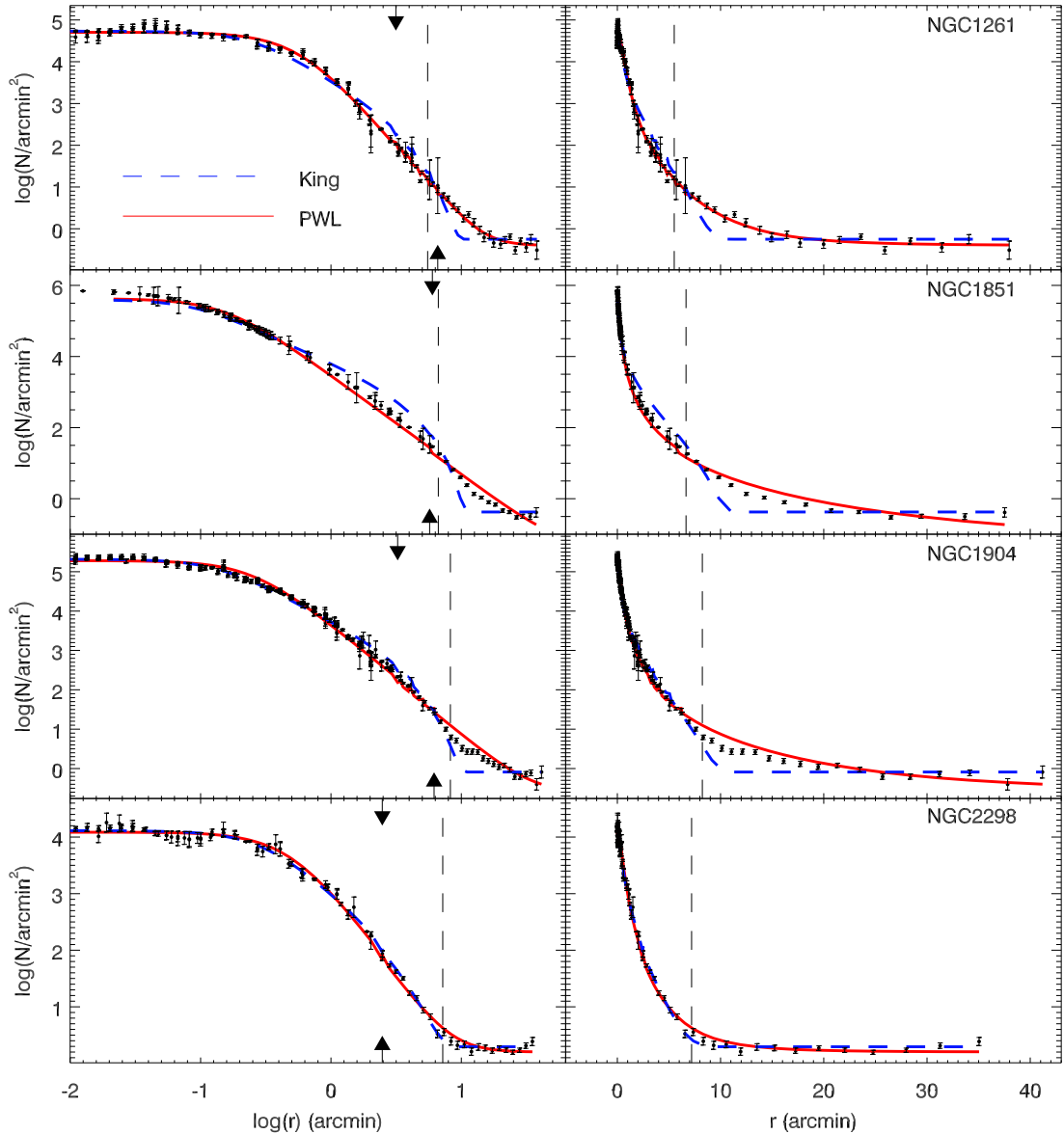}
\caption{\small{Number density profiles for NGC\,1261, NGC\,1851, NGC\,1904 and 
NGC\,2298. Blue dashed line corresponds with the best King model fitting while 
red line shows the best power-law fitting following the expression described in 
Section 3. Upper arrow shows the initial data point we have obtained from our 
data while lower arrow represents the last TGK95 data point. Vertical dashed 
line indicates the tidal radius obtained by MvM05. Note that the vertical range of the plots might differ from cluster to cluster.}}
\label{fig:prof1}
\end{centering}
\end{figure*}

\begin{figure*}
\begin{centering}
\includegraphics[scale=1.3]{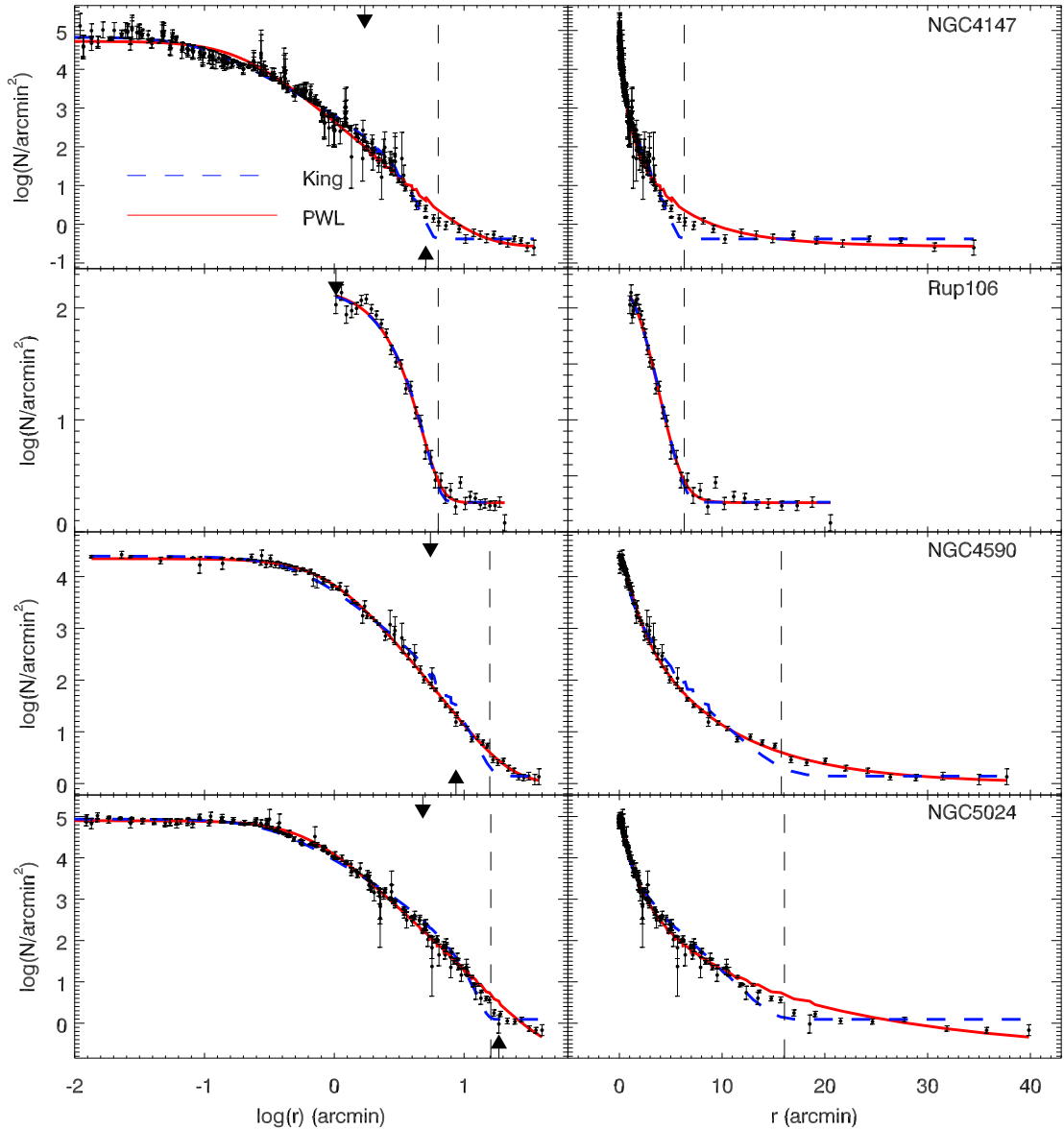}
\caption{\small{Same as Fig. \ref{fig:prof1} for NGC\,4147, Rup\,106, NGC\,4590 and
NGC\,5024.}}
\label{fig:prof2}
\end{centering}
\end{figure*}

\begin{figure*}
\begin{centering}
\includegraphics[scale=1.3]{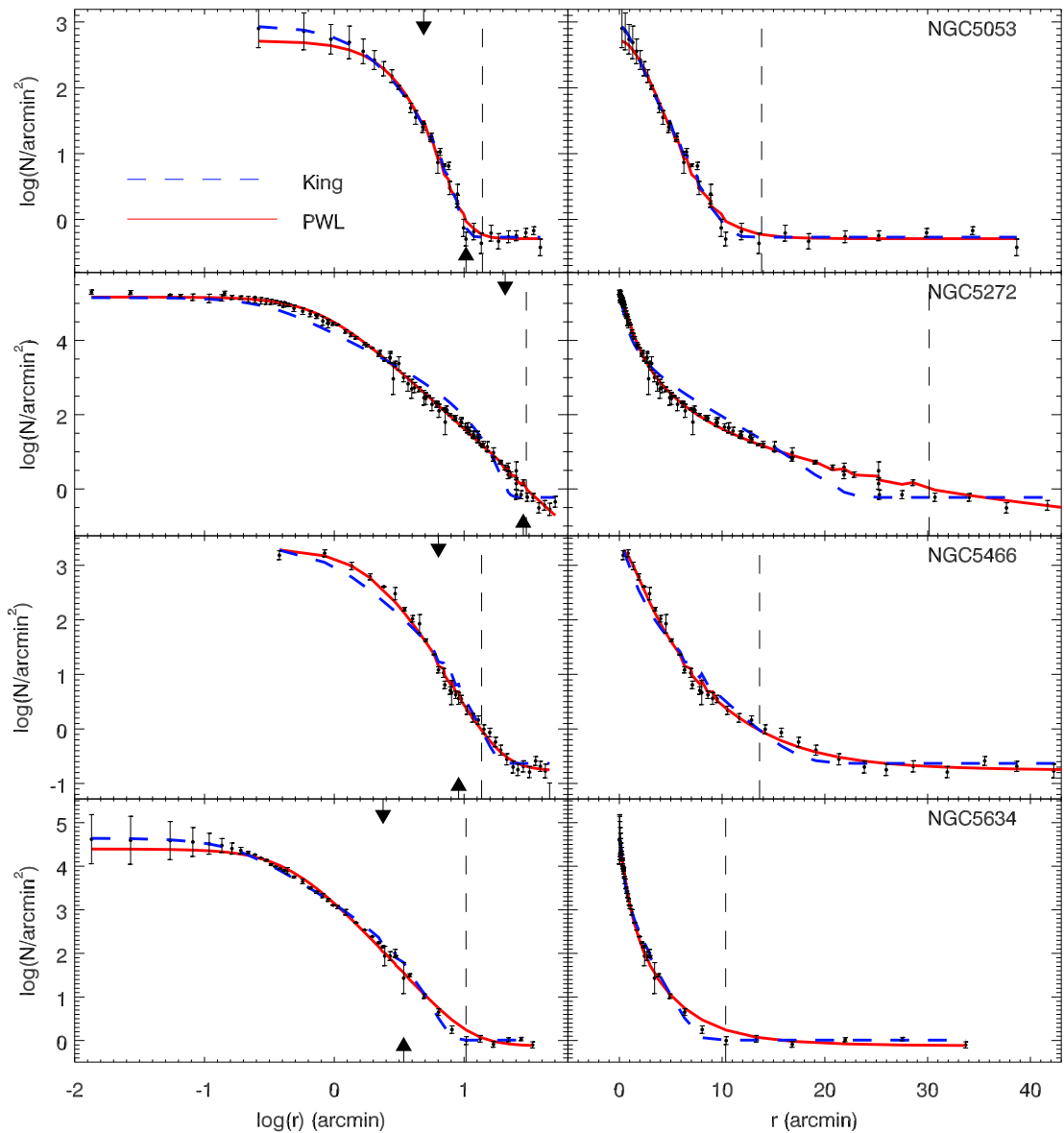}
\caption{\small{Same as Fig. \ref{fig:prof1} for NGC\,5053, NGC\,5272, NGC\,5466 and
NGC\,5634.}}
\label{fig:prof3}
\end{centering}
\end{figure*}

\begin{figure*}
\begin{centering}
\includegraphics[scale=1.3]{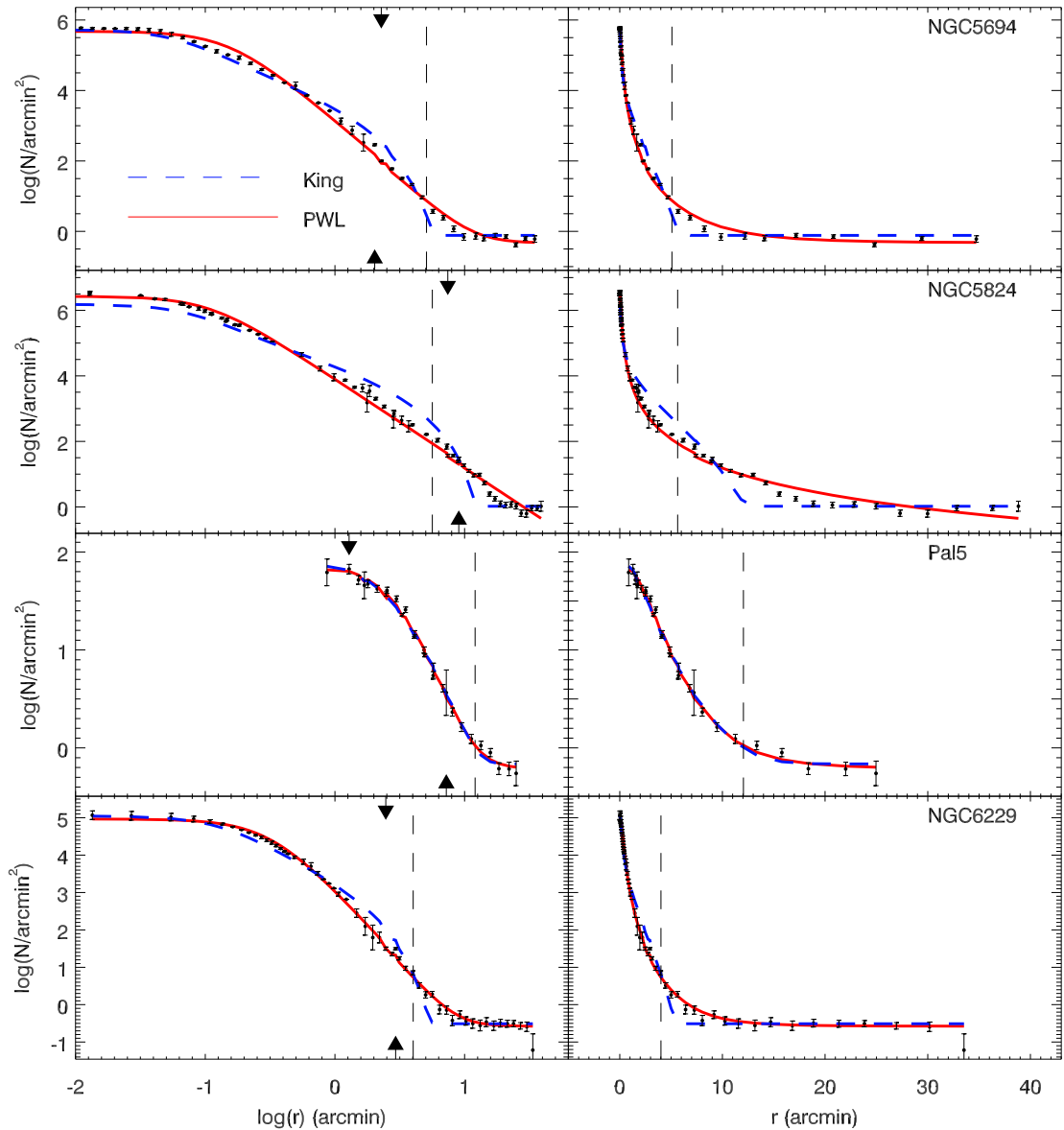}
\caption{\small{Same as Fig. \ref{fig:prof1} for NGC\,5694, NGC\,5824, Pal\,5 and
NGC\,6229.}}
\label{fig:prof4}
\end{centering}
\end{figure*}

\begin{figure*}
\begin{centering}
\includegraphics[scale=1.3]{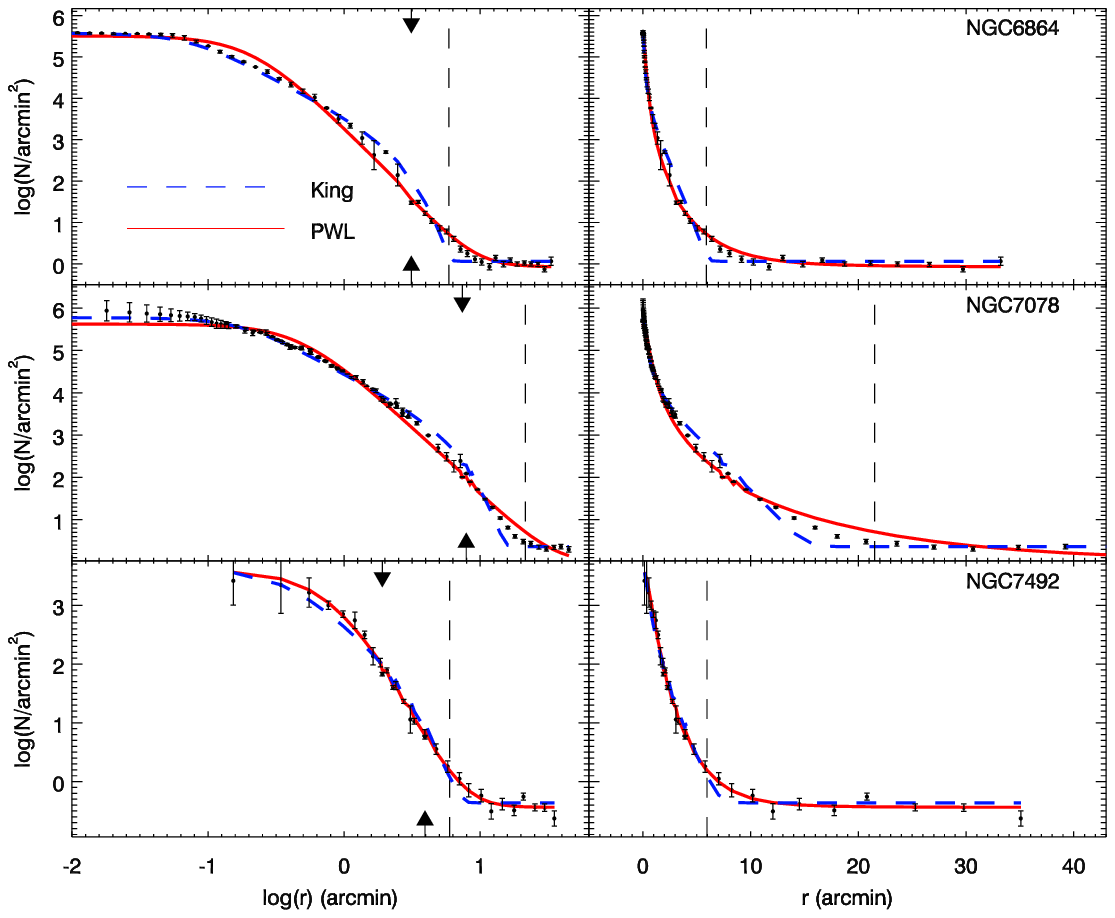}
\caption{\small{Same as Fig. \ref{fig:prof1} for NGC\,6864, NGC\,7078 and NGC\,7492
.}}
\label{fig:prof5}
\end{centering}
\end{figure*}

The derived number density profiles are shown in Figures \ref{fig:prof1} to
\ref{fig:prof5}. In order to 
obtain a better vision of the divergences between the King (blue dashed line) and 
power-law (red solid line) fittings, we have included the profiles both in 
logarithmic and linear scales in the left and right panel, respectively. 
We have also indicated the start point of our data with an arrow pointing down, whereas upward pointing arrow indicates 
the last data point of the TGK95 profile 
for each cluster (with the exception of Rup\,106 which is not contained in that 
study). 

The tidal radii previously estimated by MvM05 for these clusters are marked
 in the profiles as vertical dashed lines. For most
of the cases, the known tidal radius does not seem to be compatible
with the profiles obtained in this work. In clear cases as
NGC\,1851, NGC\,5824 or NGC\,6229, the distance at which the
truncation on the profile is expected using the TGK95 profiles alone
seems to be located within the continuous power-law profile found in our data 
set. This fact shows the risk of
computing these parameters from FOV-limited photometry.

Since in our power-law model $n_{\rm cl}(r) \propto r^{-\gamma}$ for large 
distances from the cluster centre, the parameter $\gamma$ represents the slope 
of the outer regions of the profiles and we can use it as an indicator of the 
overall shape of the profile. To have a better vision of the variation of the 
number density profiles in connection with the physical properties contained in 
Table 2, they have been sorted in ascending order of $\gamma$.

It is visible both in the density profiles and the information shown in 
Table\,2, that the tidally affected clusters NGC\,7492, NGC\,5466, Pal\,5, 
NGC\,5053 and Rup\,106 are the low-density subsample with steeper profiles. 
The effects that variable tides have in the overall structure of these clusters is 
evident with the appearance of breaks in the outer parts profiles in some of the clusters, with NGC\,5466 and Pal\,5 as best examples.  
We also know of the existence of more or less coherent tidal tails emerging from NGC\,5053 
\citep{2006ApJ...651L..33L}, NGC\,5466 \citep{2006ApJ...637L..29B}, Pal\,5 
\citep{2001ApJ...548L.165O} and NGC\,7492 \citep{Leon2000}. In this context, the shallower profile observed in the 
outer regions of NGC\,5466, Pal\,5 and possibly NGC\,7492 is generated by the dramatic impact that tidal stripping has in populating the outskirts of the cluster. Given that the second slope beyond the break only modifies slightly the global profile, we proceed with the 
analysis of these clusters in the same conditions as the rest of the sample.

On the other hand, massive clusters and more generally denser systems, present 
a relatively flatter profile with a remarkable continuous power-law distribution. 
This group coincides with what we here refer to as tidally unaffected clusters 
and in these cases we do not detect any signal of tidal truncation on the 
profiles or the existence of potential tidal debris as in the low-density group. 

Using the $\gamma$ parameter as an indicator of the overall structure, 
we conclude that, as a general rule, tidally affected clusters present $\gamma>$ 4 
while the tidally unaffected group is confined to the range 2.5 $<\gamma<$ 4. This agrees with the profile obtained for an isolated M\,31 GC by \cite{2010MNRAS.401..533M}, which ranges from $\gamma$= -2.5 in the inner parts to $\gamma$= -3.5 far from the cluster centre. 
It is also interesting to note the coincidence with the result presented by 
\cite{2010MNRAS.401.1832B}, where the existence of two populations in the Galactic GCs was found, 
based on the ratio of their half-mass to Jacobi radii. Our tidally affected (tidally unaffected) clusters are 
referred to as tidally filling (compact) in that work.

\begin{figure}
 \begin{centering}
 \includegraphics[scale=0.9]{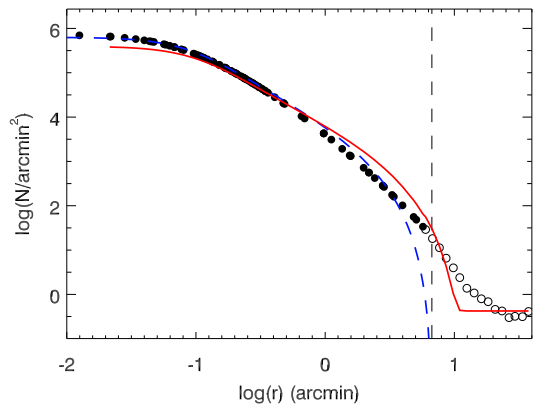}
 \caption{\small{King model fit for NGC\,1851. The fit made using only TGK95 data
 (black dots) and the combined sample constructed adding the densities measured
 in this work (open dots) are shown with red solid and blue dashed lines respectively.
Vertical line indicates the position of the edge radius derived by MvM05.}}
 \label{fig:king}
 \end{centering}
 \end{figure}

We want to remark the problems that King model has to describe the profiles
 of NGC\,1851, NGC\,5272, NGC\,5824 and NGC\,7078, four of the most 
massive clusters included in our sample, as well as NGC\,5466, NGC\,5694 or NGC\,6229. In the case of 
NGC\,7078 it is not surprising since (as NGC\,1904) is classified as 
core-collapsed GC in the  \cite{1986ApJ...305L..61D}
catalogue. In all the other cases, the 
inclusion of our data in the outermost part of the clusters drives the fit
towards small core radii and the failure of the King model in describing the overall cluster structure (see Figure
\ref{fig:king}). On the other hand, 
we find NGC\,5634 that is clearly difficult to fit with a single power-law.

Structural parameters derived from the fitting as well the masses and half-mass densities inferred from our profiles (using a power-law description) have been included in Table 2. 
The uncertainties both in $W_0$ and $\gamma$ show how important is to have the 
inner regions to constraint the overall results: in the extreme case, Rup\,106,
 the lack of information of the central parts of the cluster generates problems 
 for both models to obtain a reasonable fit.

\begin{table*}
\centering
\begin{tabular}{lcccrclllrccll}
\hline
\multicolumn{4}{l}{} & \multicolumn{5}{c}{Power law} &\multicolumn{5}{c}{King} \\
Cluster& $\log(M)$ & $\log(\rhoh)$ & $\log(\trh)$ & 
$\gamma$ & $r_{\rm 0}$ & $\reff$ & $\chi^{2}_{\rm TGK95}$ &   $\chi^{2}_{\rm New}$  & 
$W_0$ & $\rc$ & $\reff$ &  $\chi^{2}_{\rm TGK95}$ &   $\chi^{2}_{\rm New}$\\

       & [$\sm$]   & [$\sm$/pc$^{3}$]  &   [yr]        &
       &  ['] &  [']  &  & & 
      & [']   & [']  & &\\		      
\hline
\hline
\\
NGC\,1261 &  5.20  &   2.14  &  9.26 &   3.68$^{+0.07}_{-0.17}$   & 0.58  & 0.67   & 114.7  &  67.1   & 7.17$^{+0.15}_{-0.06}$  &   0.29 & 0.92     & 416.6 & 449.9       \\
NGC\,1851 &  5.68  &   3.52  &  8.99 &    2.69$^{+0.05}_{-0.06}$  & 0.16  & 0.54   & 131.2  &  101.6  &  8.66$^{+0.17}_{-0.09}$ &   0.11 & 0.96   & 221.2 & 164.4     \\
NGC\,1904 &  5.34  &   2.64  &  9.11 &   2.71$^{+0.05}_{-0.04}$   & 0.26  & 0.68   & 475.5  &  165.9  & 8.06$^{+0.07}_{-0.08}$  &   0.16 & 0.89     & 343.8 & 317.2     \\
NGC\,2298 &  4.59  &   1.91  &  8.82 &   3.08$^{+0.30}_{-0.10}$   & 0.50  & 0.65   & 75.3   &  32.4   & 6.89$^{+0.15}_{-0.24}$  &   0.32 & 0.90     & 65.0  & 20.2     \\
NGC\,4147 &  4.88  &   2.12  &  8.97 &   2.81$^{+0.07}_{-0.05}$   & 0.19  & 0.53   & 781.3  &  110.7  &7.96$^{+0.06}_{-0.12}$  &   0.10 & 0.52      & 376.7 & 232.7    \\
Rup\,106  &  4.74  &   0.06  & 9.89  &   15.28$^{+4.10}_{-6.98}$  & 6.56  & 2.18   & -      &	28.5      & 0.51$^{+2.43}_{-0.14}$  &   6.24 & 2.40     & -     &  79.4  \\
NGC\,4590 &  5.02  &   1.73  &  9.33 &   3.15$^{+0.18}_{-0.16}$   & 0.95  & 0.72   & 49.6   &  15.3   & 7.18$^{+0.18}_{-0.13}$  &   0.56 & 1.80     & 112.3 & 51.5   \\
NGC\,5024 &  5.77  &   2.06  &  9.81 &   2.98$^{+0.06}_{-0.07}$   & 0.63  & 0.74   & 262.8  &  89.0   & 7.56$^{+0.09}_{-0.06}$  &   0.36 & 1.44     & 491.6 & 110.3    \\
NGC\,5053 &  5.07  &   0.17  &  9.75 &   7.62$^{+3.96}_{-1.28}$   & 4.44  & 2.35   & 28.9   &  13.1   & 4.36$^{+0.82}_{-0.97}$  &   1.62 & 2.09     & 15.6  & 7.0	      \\
NGC\,5272 &  5.78  &   2.72  &  9.48 &    3.18$^{+0.08}_{-0.05}$  & 0.77  & 0.87   & 95.9   &  36.3   & 8.06$^{+0.12}_{-0.10}$ &   0.36 & 1.99  & 441.4 & 62.6	      \\
NGC\,5466 &  4.65  &   0.51  &  9.56 &    4.05$^{+0.43}_{-0.27}$  & 2.06  & 2.03   & 14.1   &  39.5   & 6.58$^{+0.37}_{-0.33}$ &   0.91 & 2.23	 & 41.9  & 49.0      \\
NGC\,5634 &  5.26  &   1.66  &  9.54 &   3.20$^{+0.28}_{-0.16}$   & 0.45  & 0.62   & 41.7   &  59.5   & 7.56$^{+0.30}_{-0.38}$  &   0.19 & 0.77     & 9.3   & 20.4	      \\
NGC\,5694 &  5.49  &   2.35  &  9.41 &   3.14$^{+0.15}_{-0.08}$   & 0.16  & 0.40   & 70.7   &  40.3   & 8.56$^{+0.19}_{-0.19}$  &   0.06 & 0.53   & 44.9  & 115.8    \\
NGC\,5824 &  5.98  &   2.76  &  9.64 &    2.62$^{+0.05}_{-0.04}$  & 0.11  & 0.46   & 100.3  &  136.6  &  9.45$^{+0.21}_{-0.12}$ &   0.08 & 1.26   & 398.0 & 269.3    \\
Pal\,5    &  4.10  &  -1.25  &  9.78&   4.34$^{+3.58}_{-0.92}$   & 3.79  & 3.41   & 1.0    &  9.3    & 4.54$^{+1.28}_{-1.36}$  &   2.43 & 3.25     & 1.7   & 15.7	      \\
NGC\,6229 &  5.49  &   2.27  &  9.46 &   3.80$^{+0.24}_{-0.09}$   & 0.32  & 0.49   & 19.0   &  36.0   & 7.36$^{+0.21}_{-0.15}$  &   0.14 & 0.49     & 112.6 & 203.0       \\
NGC\,6864 &  5.75  &   3.11  &  9.28 &   3.31$^{+0.16}_{-0.08}$   & 0.22  & 0.46   & 124.5  &  40.3   & 8.16$^{+0.20}_{-0.17}$  &   0.09 & 0.54     & 26.4  & 184.3       \\
NGC\,7078 &  6.14  &   3.40  &  9.48 &   2.96$^{+0.10}_{-0.07}$   & 0.48  & 0.77   & 181.4  &  54.1   & 8.26$^{+0.09}_{-0.20}$  &   0.21 & 1.39     & 129.5 & 109.8    \\
NGC\,7492 &  4.42  &   0.66  &  9.30 &   3.94$^{+0.56}_{-0.27}$   & 0.81  & 0.83   & 10.2   &  16.9   & 6.43$^{+0.73}_{-0.62}$  &   0.39 & 0.97     & 43.3  & 54.9	      \\
\\									     
\hline
\hline
	  \end{tabular}

\caption{\small{Physical and structural parameters. Masses,densities and 
half-mass radius relaxation times derived as described in Section 2 and using
 the new $\rh$ estimations obtained from our best power-law fitting. Ranges of confidence 
 for $W_0$ and $\gamma$ have been included to show the discrepeances in some of 
 the clusters, specially Rup\,160. } }

\end{table*}

\section{Discussion}

\subsection{Power-law template vs. King model}

The number density profiles presented in this paper show that, in most of 
the cases, the outer parts of the clusters extend to large
distances from the centre, generally beyond the edge radius inferred 
from the previous studies based on datasets restricted to the innermost region. 
As a consequence, the systematic application of the King model to previous 
data and the lack of complete radial profiles seems to have generated the idea 
that these profiles are truncated, whereas here we show that in some cases the 
profile is well described by a continuous power-law down 
to our detection limit.

\begin{figure*}
\begin{centering}
\includegraphics[scale=0.9]{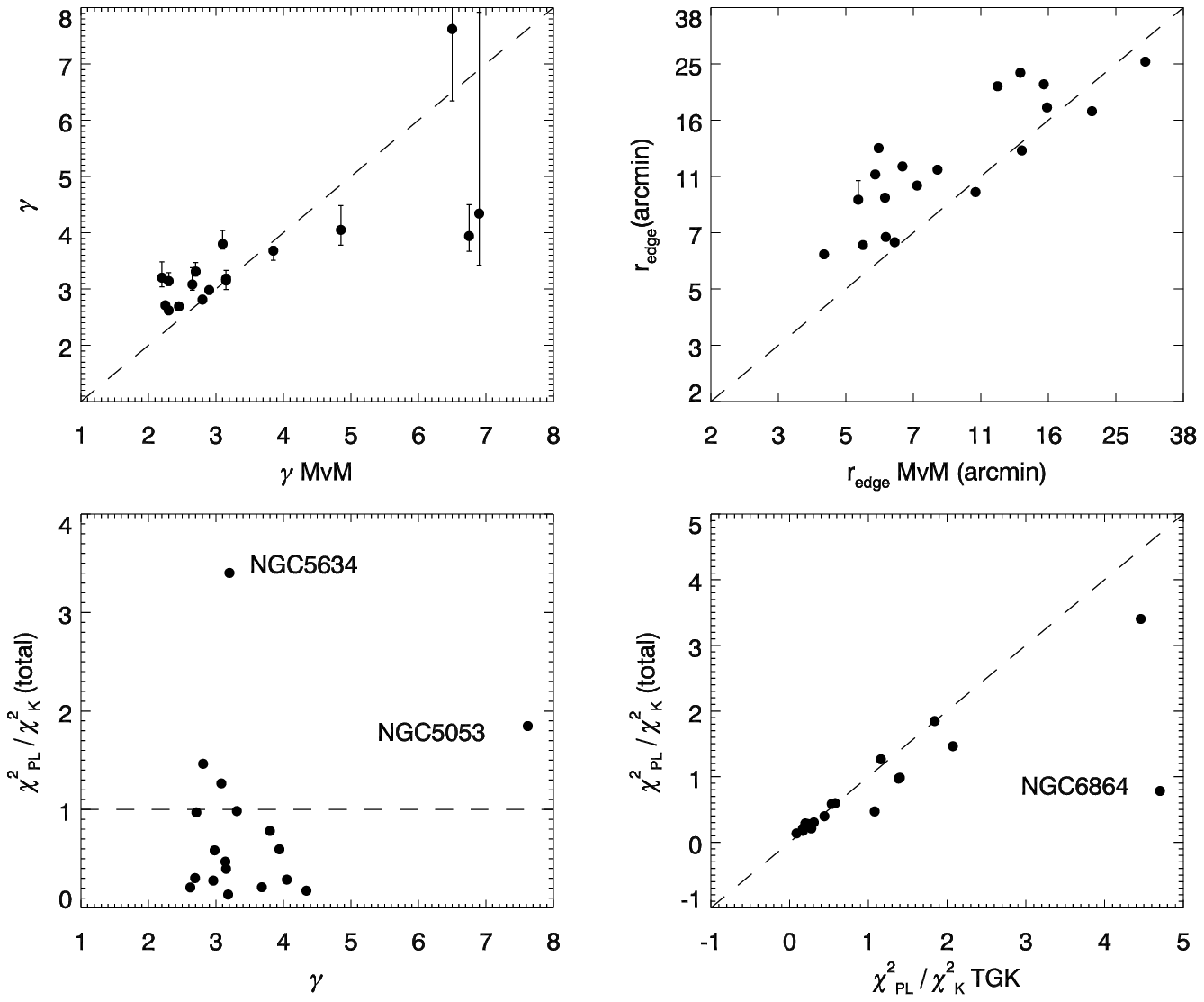}
\caption{\small{Comparison between our $\gamma$ (upper-left panel) and $\rt$
(upper-right panel) values
  with those derived by MvM05. In
  bottom panels we show the $\chi^{2}_{\rm PL}/\chi^{2}_{\rm K}$ ratio is shown
  as a function of $\gamma$ (bottom-left panel) and the same ratio measured on
  TGK95 data (bottom-right panel).}}
\label{fig:comp}
\end{centering}
\end{figure*}

To analyse the effect that increasing the observed FOV has on the derived 
properties of GCs, we have plotted our best fit $\gamma$ and $\rt$ 
against those derived by MvM05\footnote{{Note that MvM05 use $\gamma$ to describe the power-law decline of the {\it 3-dimensional} number density profile. We, therefore, subtracted 1 from their values to be able to compare to our $\gamma$, that describes the decline of the number density profile {\it in projection}, i.e. $\gamma_{\rm 3D} = \gamma_{\rm 2D} + 1$.}}
(which are based on the TGK95
surface brightness profiles) in Figure \ref{fig:comp} (upper panels). From this comparison, 
the presence of clear trends is evident: for tidally unaffected GCs 
($\gamma\leq 4$), we find slightly steeper profiles (larger $\gamma$), while for the tidally affected clusters 
NGC\,5466, Pal\,5 and NGC\,7492 the profiles are slightly shallower (smaller $\gamma$). This result is 
expected because 
of the presence of tidal debris in the outermost parts of these clusters which 
were not contained in the original TGK95 data. 
New edge radii are also remarkably bigger than the previous values reported 
by MvM05 with a mean variation of $\sim$+\,40 $\%$, excluding the 4 GCs which present smaller derived edge radii and those which are poorly fitted by King models (see below). With these new results, we do not detect signatures, at least in the tidally
 unaffected group, of truncation on the profiles. 

To investigate the suitability of the King model and a power-law to describe 
our GCs, we compared the computed $\chi^{2}$. In Figure \ref{fig:comp} we plot the 
rate of the $\chi^{2}$ of the power-law fit, $\chi^{2}_{\rm PL}$, over the $\chi^{2}$ following from the King model fit, $\chi^{2}_{\rm K}$,
 as a function of $\gamma$ (bottom-left panel) and 
the same ratio if we only include 
the TGK95 data (bottom-right panel). According to these results, 
the power-law template is a better approximation to the observed number density 
profiles for $\sim$ 2/3 of the clusters in our sample with the 
remarkable exceptions of NGC\,5634 with 
$\chi^{2}_{\rm PL}/\chi^{2}_{\rm K} \ge$ 3. Again, the density profiles of the tidally 
affected clusters (with the exception of NGC\,5053) are better reproduced by power-law fits.
The same result can be found also using only the inner part of the profile from TGK95. 
It is interesting to note the striking difference in the
$\chi^{2}_{\rm PL}/\chi^{2}_{\rm K}$ ratio of NGC\,6864 (member of the
tidally unaffected group of clusters) which is better fitted by a King model when
only the inner region is considered, while showing a power-law shape when the
entire profile is considered.

The reason for the deviation from the King model profile in the group of 
tidally affected GCs can be found in
the working hypothesis made in \cite{1966AJ.....71...64K} models on the radial truncation 
at the zero-energy surface. In fact, while stars with energy above the critical energy (i.e. a
velocity larger than the local escape velocity) are actually expected to 
evaporate from the cluster, in real clusters they remain marginally bound to the
system for a timescale comparable to the cluster orbital period around the
parent galaxy \citep{1987ApJ...322..123L}, or even longer \citep{2000MNRAS.318..753F}.
This effect produces a significant overdensity of
stars in the outermost regions of the cluster with respect to the prediction of
a King model without affecting the inner profile \citep[see also][]{1999MNRAS.302..771J}. 
This effect is more evident in clusters in the evaporation regime (i.e. tidally
affected) and 
those subject to a stronger interaction with the external tidal field.
The assumption of the King edge radius as a real physical limit of the cluster, 
leads to the use of the term 'extra-tidal stars' for those cluster members 
outside this radius. This has been confirmed in several GCs during the 
last years, with NGC\,1851 \citep{2009AJ....138.1570O}, IC\,4499 
\citep{2011arXiv1103.4144W} or NGC\,5694 \citep{2011arXiv1105.2001C} as recent examples, 
and can be noticed also in an 
important part of our profiles (see Figures \ref{fig:prof1} to \ref{fig:prof5}).
Of course, despite of the above limitations, King models remain a valid
representation of a GC (at least in the inner part of the profile). 
However, because the power-law template is a reasonable choice to describe the derived number density profiles, 
we decided to use the outer slope $\gamma$ as an indicator of the overall structure of our target GCs in the
following sections. 
The main consequence of assuming a power-law
as the best template to model a radial profile is that GCs
have an infinite extension. Of course we assume that at some point there exists 
a physical limit at the first Lagrangian point or Jacobi radius.

In the following we analyse the 
impact that both external and internal factors have in the observed values of 
$\gamma$. 

\subsection{External factors}

It is well known that the structure of a GC is influenced by the interaction with 
the host Galactic potential \citep{1972ApJ...176L..51O,1999ApJ...513..626G,1999MNRAS.302..771J}. 
In fact, at every passage across the Galactic disc
and at every pericentric passage, 
a transfer of kinetic energy occurs in the form of compressive (disc) and
tidal (bulge) shocks. As a result, the escape of high velocity stars is
triggered while the overall energetic budget of the cluster
is continuously altered. The theory of the structural evolution of GCs as a
result of external tidal effects has been extensively investigated with
analytical approximations \citep{1987ApJ...322..113C}, through Fokker-Planck 
modelling \citep{1999ApJ...513..626G} and $N$-body simulations \citep{2003MNRAS.340..227B,2008MNRAS.389L..28G}.

It is therefore interesting to check how the strength of these external factors
correlates with the shape parameters of our sample of clusters.
This task requires the knowledge of the cluster orbits to quantify the impact of
the Galactic tidal field.
Unfortunately, among the 19 clusters of our sample, only 10 of them have 
known proper motions so that orbital parameters can be derived only for this
subsample. 

\begin{figure}
 \begin{centering}
 \includegraphics[scale=0.9]{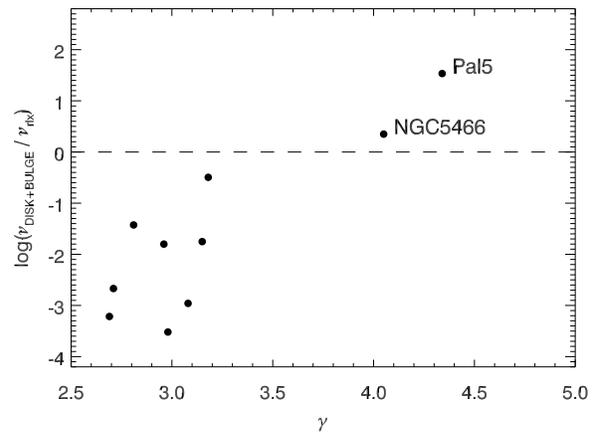}
 \caption{\small{Ratio between the disc+bulge shocking (from Allen et al. 2006) and
 relaxation from (Baumgardt \& Makino 2003) destruction rates as a function of 
 $\gamma$. The two clusters belonging to the tidally
 affected group (Pal\,5 and NGC\,5466) are indicated. }}
 \label{fig:relative}
 \end{centering}
 \end{figure}

\begin{figure*}
\begin{centering}
\includegraphics[scale=1]{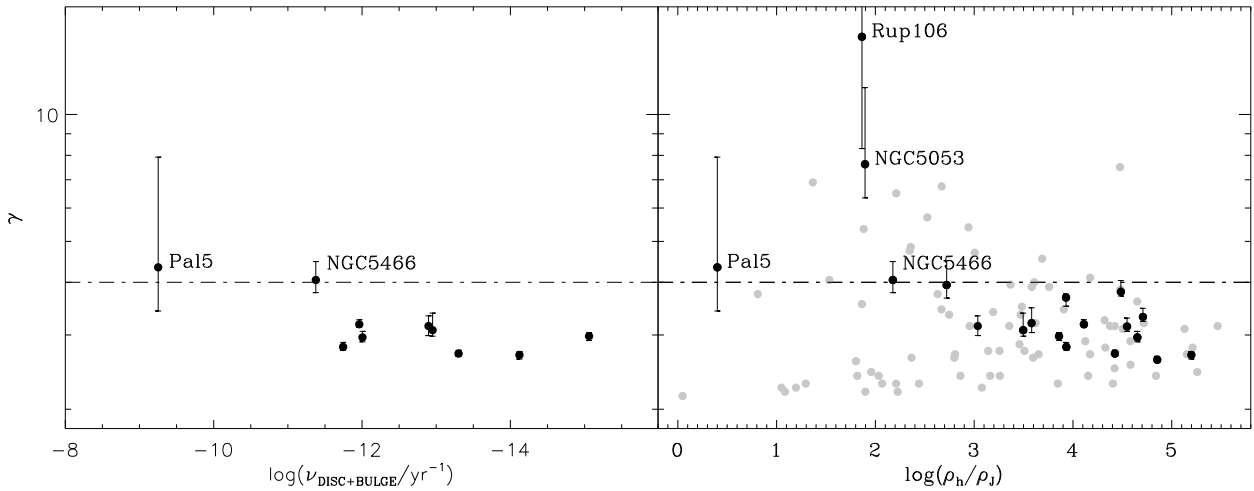}
\caption{\small{Left: $\gamma$ parameter as a function of the disc+bulge shocking
destruction rate $\nu$ from Allen et al. (2006). The two clusters belonging to the tidally
 affected group (Pal\,5 and NGC\,5466) are indicated. Right: variation of $\gamma$ as a 
 function of $\rho_{\rm h}$/$\rho_{\rm J}$. The dashed line in both panels 
 indicates 
the tentative border between the two categories of clusters (at $\gamma=$ 4). 
Grey points mark the sample of Galactic GCs studied by MvM05.}}
\label{fig:orb}
\end{centering}
\end{figure*}

As a first test we estimated the relative importance of the destruction rate 
due to relaxation and shocks. For this purpose we adopted the destruction
rates due to disc + bulge shocking computed by \cite{2006ApJ...652.1150A} and the contribution 
of relaxation taken from the $N$-body simulations of \cite{2003MNRAS.340..227B}\footnote{As 
noticed by  \cite{1997ApJ...474..223G}, the
combined effect of shocks and relaxation is generally non-linear. Therefore, 
although a rigorous decoupling of these two effects is not possible, such an 
approach must be considered an approximated way to quantify the relative 
importance of the two effects.}. In Figure \ref{fig:relative} the ratio $\nu_{\rm disc + bulge}/\nu_{\rm rlx}$ is
shown for the 10 GCs of our samples with known orbits.
For two clusters shocks turn out to be important (Pal\,5 and NGC\,5466), 
both belonging to the tidally affected group. For the other GCs (all
belonging to the tidally unaffected group) evaporation due to relaxation is the 
dominant effect. The position of these clusters in the $\gamma-\nu_{\rm disc + bulge}$ plane 
is shown in the left panel of Figure \ref{fig:orb}. 
It is evident that the two tidally affected clusters, characterized by steeper
profiles, are subject to significantly stronger shocks with respect to the
group of tidally unaffected clusters.

\nocite{2003MNRAS.340..227B}

In absence of accurate orbital parameters it is not possible to check the above
correlations for the entire sample of GCs.
However, as a general rule, it is expected that the interaction between low-density 
clusters and the 
Galaxy become more important because their densities are close to the background density of the Galaxy. In the right 
panel of Figure \ref{fig:orb} we compare the half-mass radius density with the density 
of field stars within the Jacobi radius calculated as 

\begin{equation}
  \begin{centering}
    \rho_{\rm J} \sim 5.376 \Big(\frac{\rg}{\rm kpc}\Big)^{-2}  M_{\odot} \rm pc^{-3}
  \end{centering}
\end{equation}

according to appendix B in \cite{G11}. As expected, the tidally affected 
clusters are restricted to the $\log(\rho_{\rm h}/\rho_{\rm J})<$ 3 region of that 
plot while the rest of $\gamma<$ 4 clusters are at least 3 orders of magnitude 
denser than $\rho_{\rm J}$. If we add to our sample those 
included in MvM05 (grey circles in Figure \ref{fig:orb}) we find that clusters in the 
range $\log(\rho_{\rm h}/\rho_{\rm J}) \ge$ 3 follow the same tendency to be 
represented by $\gamma<$ 4. On the other hand, there is a group of objects at 
$\log(\rho_{\rm h}/\rho_{\rm J}) \sim$ 1 and close to $\gamma$ = 2 which appear 
to stray from this behaviour. We will discuss the position of this group of 
clusters in the next section.

\begin{figure*}
 \begin{centering}
 \includegraphics[scale=1]{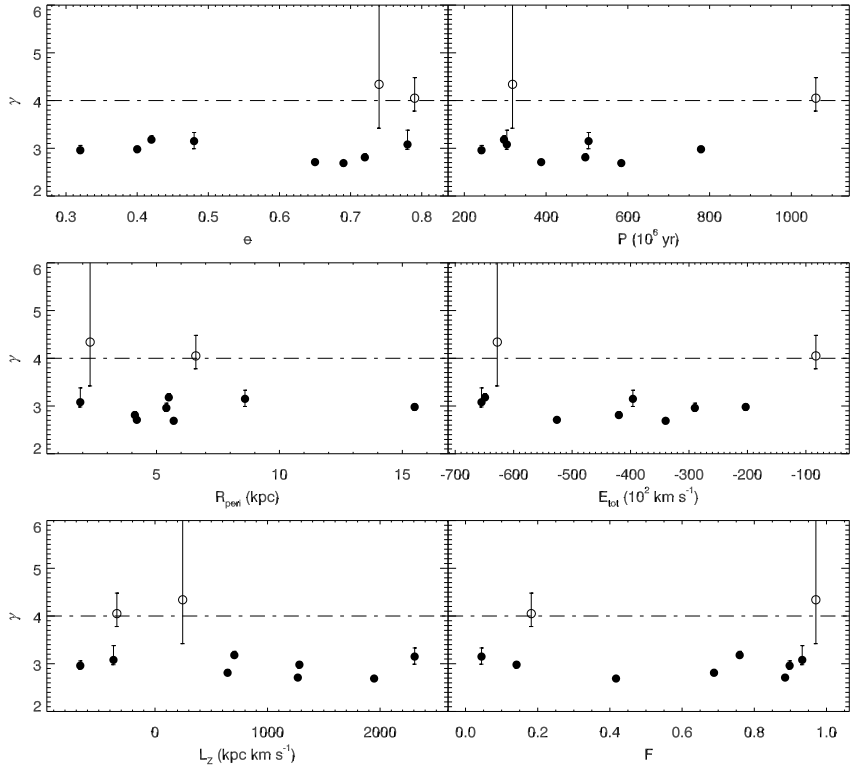}
 \caption{\small{$\gamma$ as a function of the orbital parameters: eccentricity
 (upper left panel), period (upper right panel), pericentric distance (middle
 left panel), energy (middle right panel), angular momentum (bottom left panel)
 and phase (bottom right panel). The location of the two clusters of the tidally
 affected group (Pal\,5 and NGC\,5466) are indicated in each panel with open circles.}}
 \label{fig:parameters}
 \end{centering}
 \end{figure*}

As a further test we also correlated the outer slope parameter $\gamma$ with 
the main cluster orbital parameters total energy, z-component of the angular momentum, pericentre radius, orbital period and eccentricity
($E, L_{\rm Z}, \rm R_{\rm peri}, P $ and $e$ respectively) taken from \cite{1999AJ....117.1792D}
and constructed an additional parameter 
F=(R$_{\rm G}-$R$_{\rm peri})$/(R$_{\rm apo}-$R$_{\rm peri}$) which is an indicator of the 
current position along the orbit of the clusters. With this definition, F=1 
when the cluster is at the apocentre while F=0 when it is at its 
pericentre. In these planes the orbits of the two tidally affected clusters are those with the
largest eccentricity and the smallest modulus of angular momentum. 
This confirms again that eccentric orbits, which make the cluster subject to a 
significant variation of the surrounding external field, can lead clusters
toward the evaporation regime. A group of tidally unaffected clusters
exibit orbits with similar eccentricities and angular momentum. It is
interesting to note that all these clusters have been suspected to be accreted \citep[e.g. NGC1851, NGC1904, NGC2298 and NGC4147; ][]{2010MNRAS.404.1203F} or are very massive (NGC7078). 
Accreted clusters might have different evolutionary histories from Galactic GCs, 
passing a significant part of their evolution in smaller 
systems as dwarf galaxies before the accretion, so this could explain why accreted and in-situ 
clusters present a different overall structure, even if they have similar 
orbital parameters. Mass represents also an important parameter since massive
clusters could resist the strong tidal interaction that lead towards high gamma 
values. As we have commented, clusters on highly-eccentric orbits as NGC\,5466 and 
Pal\,5 are influenced by the interaction with the Milky Way significantly, 
leading to the formation of an external  (and different) power-law distribution. 
No other significant trends are visible in the plots shown in Figure \ref{fig:parameters} between $\gamma$ and the
other orbital parameters.

\subsection{Internal evolution}

 \begin{figure*}
 \begin{centering}
 \includegraphics[scale=1]{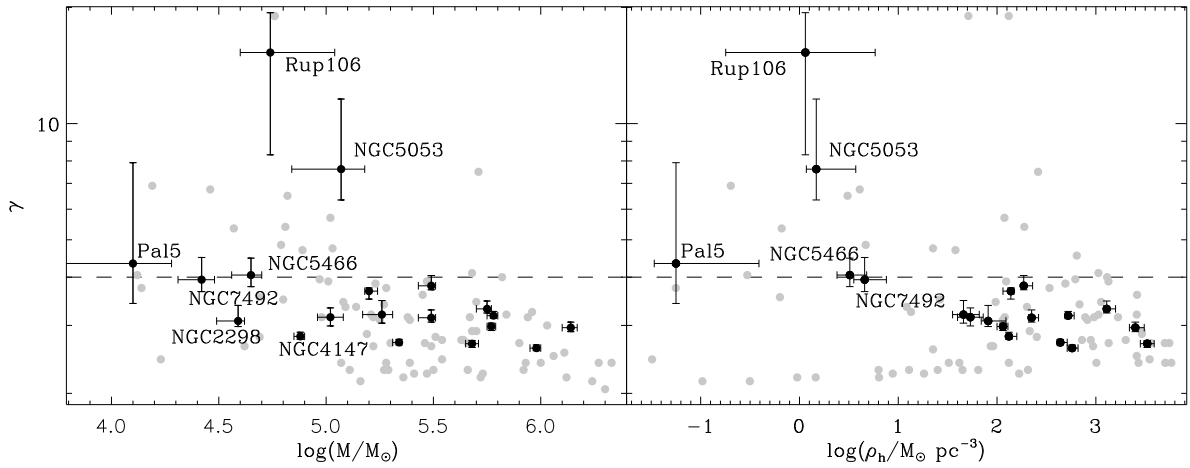}
 \caption{\small{$\gamma$ as a function of mass and density. 
 The dashed line in both panels indicates 
the tentative border between the two categories of clusters (at $\gamma=$ 4).
 Grey points mark the sample of Galactic GCs studied by MvM05. }}
 \label{fig:dens}
 \end{centering}
 \end{figure*}

After evaluating the influence of external factors on the observed number 
density profiles, we looked at the correlations between the general physical 
parameters of the cluster with the slope $\gamma$. As in stellar physics, the 
global mass is 
always one of the first suspects to drive the behaviour of the structure of a 
GC and we have tested this correlation in the left panel of Figure \ref{fig:dens}. 

It has been already noticed (Section 4) that all the clusters of the tidally unaffected 
subsample, formed by massive clusters, present flatter radial profiles while the
less massive tidally affected clusters have large values of $\gamma$. 
A similar 
relation was found by \cite{2000ApJ...539..618M} using in that case the 
concentration $\log(\rt/\rc)$ as a function of the cluster luminosity.  

A possible interpretation of this behaviour can be searched by looking at the
role of binaries in the long-term evolution of GCs.
Low-mass clusters are able to form and
retain more binaries than massive clusters where they are more efficiently 
destroyed in single-binary and binary-binary close encounters 
\citep{1994MNRAS.269..241G,2008MNRAS.388..307S,2009ApJ...707.1533F}.
While the natural evolution of GCs is toward high-concentration structures \citep{1961AnAp...24..369H}
 a large fraction of binaries prevent the contraction of the core and maintain the
cluster in a quasi-steady state of binary burning where the continuous loss of
energy due to evaporation is balanced by the energy heating of binaries 
\citep{1991ApJ...370..567G,1998MNRAS.299.1019V,2007ApJ...658.1047F}. 
Unfortunately, the binary fraction has only been estimated for NGC 4590
(tidally unaffected), NGC 5053 and NGC 5466 (tidally affected) 
\cite[see ][]{2007MNRAS.380..781S} and appear to cover the same 
range ($9.5<\xi<14.2\%$) within the errors. Although the available data do not
allow to confirm the scenario proposed above, other studies have suggested an 
anti-correlation between the mass and the fraction of binary systems in larger samples of Galactic GCs 
\citep{2008MmSAI..79..623M,2008MNRAS.388..307S,2010MNRAS.401..577S}. In addition,
mass loss due to stellar evolution drives the dynamical evolution in early
stages and causes
low-mass stellar systems to expand to their tidal boundary faster than their
more massive counter-parts \citep{2010MNRAS.408L..16G}. This, together with the
external factors discussed in the previous section, 
could explain the tendency found in the low-mass tidally affected clusters to 
have prominent cores, large half-mass radii and larger values of $\gamma$.   

We have to analyse the exception of NGC\,2298 and NGC\,4147 in the $\gamma$ vs. 
$\log(M)$ plot. Despite their low masses, these clusters are included in the 
tidally unaffected group according to Table 2. Special birth conditions or an 
external origin could explain these differences. This last hypothesis has been 
widely supported for NGC\,4147 \citep{2003AJ....125..188B, 
2010MNRAS.404.1203F} since it lies in the projected 
position path of the Sagittarius tidal stream sharing the same energy and
angular momentum. Similarly, NGC\,2298 has been associated to 
the controversial Canis Major dwarf galaxy \citep{2004MNRAS.348...12M,2010MNRAS.404.1203F} 
although it presents a retrograde orbit incompatible with that of this 
over-density but also with the general rotational pattern of the Milky Way.
In this picture, GCs formed in less denser systems, might live as isolated 
(tidally unaffected) clusters before the accretion by the Milky Way.

The derived outer slope $\gamma$ vs. the half-mass density $\rho_{\rm h}$ is
shown on the right panel of Figure \ref{fig:dens}. 
Also in this case, a striking correlation is evident in the sense that denser
clusters have on average flatter profiles. In this case, denser clusters are 
both expected to be more resistant to the external tidal stress exerted by the 
Milky Way (see Section 5.2) and dominated by the effect of relaxation (eq.
\ref{eq:trh}).
In Figure \ref{fig:dens} we have also included the sample of clusters studied 
by MvM05. The trend found in the relations 
between mass, half mass density and $\gamma$ are also visible here although the 
presence of a branch of massive clusters with $\gamma \sim$ 2 and reaching very 
low densities decreases the significance of the $\gamma-\rho_{\rm h}$ correlation. 

Of course, the MvM05 sample contains an order of magnitude larger number of
objects and constitutes a more robust statistical sample to check such correlations.
However, it is noticeable that for most of these object the TGK95 
profiles cover a very limited FOV (including only the first central arcminutes).
As shown in Section 4, an incomplete radial coverage can significantly 
alter the estimate of the profile slope leading to a more uncertain determination
of both $\gamma$ and $\rho_{\rm h}$.
It is therefore possible that the surprisingly low half-mass density of these
object is due to an underestimate of $\gamma$ and, consequently, $\rho_{\rm h}$.   

\begin{figure}
 \begin{centering}
 \includegraphics[scale=1]{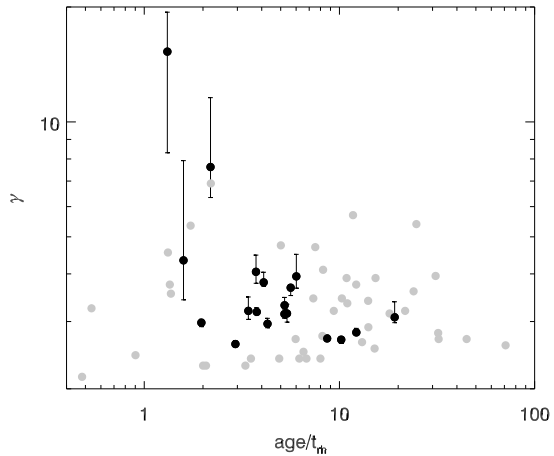}
 \caption{\small{$\gamma$ as a function of $age/t_{rh}$. Grey dots mark the
 Galactic GCs of the MvdM sample.}}
 \label{fig:ages}
 \end{centering}
 \end{figure}

Connecting the structure of our sample of Galactic GCs with their dynamical age 
is not simple because of their narrow range of relative ages 
\citep{2009ApJ...694.1498M}. In Figure \ref{fig:ages}, we show the parameter $\gamma$ as a 
function of the ratio $age/t_{\rm rh}$. 
On average, clusters with lower ratios of ${\rm age}/\trh$
have larger values of $\gamma$ even if the large scatter prevents from any firm
conclusion on the existence of any reliable correlation. The same behaviour is 
visible among the clusters in the MvdM05 sample with the exception of the group of globulars
 with $\gamma \sim$ 2 already mentioned above which show a lower $age/\trh$. It is difficult to say something about the relation between $\gamma$ and the dynamical age of clusters. This is firstly because clusters in the tidally unaffected regime have expanded such that their relaxation times have become a fixed fraction of their age, roughly equal to $\sim1/10$ for clusters of equal masses \citep{1965AnAp...28...62H, 1984ApJ...280..298G} and $\sim1/3$ for clusters with a globular type stellar mass-function \citep{G11}. The majority of clusters in our sample is in this regime and forms a cloud of points in Fig.~\ref{fig:ages}, Secondly, the evolution of clusters with an age roughly equal to $\trh$ is probably driven by external factors due to their low density, hence $\trh$ is not telling us much about the evolution for these objects.

\section{Conclusions}

We have presented the number density profiles of 19 Galactic GCs obtained with 
deep wide-field photometry. 
The comparison of our results, fitting both King models and power-law templates,
with those derived by previous studies shows the importance that data with a 
wider FOVs have on our understanding of these stellar systems.
In particular, the edge radii estimated from this analysis have been found to
be $\sim20\%$ larger than what found in previous works based on shallower
surface brightness profile.
The number density profiles of our sample of GCs can be reasonably fitted by
both King models and power-law templates. The latter ones appear to be a better
representation for $\sim 2/3$ of the observed profiles including some of those clusters showing evidences of
tidal tails \citep[Pal\,5 and NGC\,5466,][]{2001ApJ...548L.165O,2006ApJ...637L..29B}, in agreement with the prediction of
$N$-body simulations\citep{2006MNRAS.367..646L,2009ApJ...698..222P,1999MNRAS.302..771J}. 

It has been found that the slope $\gamma$ is linked
to the membership of the cluster to the group of "tidally affected" or "tidally
unaffected" clusters (defined by \cite{G11} on the basis of the
position in the $\rho_{\rm h}(M,R_{\rm G})$ plane). 
The group of tidally affected clusters, constituted mainly of low-mass objects
with prominent cores, is characterized by steeper profiles ($\gamma>$ 4) while
tidally unaffected clusters show flatter profiles extending to large distances
from their compact cores.

We investigated the dependence of the outer slope of the density profile 
$\gamma$ on the internal structural evolution of the cluster (based on
the relaxation process) and on external factors (i.e. tidal 
shocks) using a subsample of 10 GCs with known orbits. 
External factors have been found to be dominant in the two tidally affected
clusters (Pal\,5 and NGC\,5466), which have preferentially more eccentric orbits 
and larger destruction rates than tidally unaffected clusters.
This finding agree with the evidence of massive tidal tails around these two
clusters (see above) indicating their strong interaction with the Galactic
potential.
For the other clusters of the sample, internal processes can be considered the 
main mechanisms of dynamical evolution.

We revealed a slight correlation between $\gamma$ and the cluster mass and half-mass
density which can be interpreted as a consequence of the fact that the most
massive and dense tidally unaffected clusters are currently in an expansion dominated 
phase \citep{G11}. The connection between slope $\gamma$ and 
half-mass density is less significant in the larger sample by MvM05 which is however
limited to the innermost fit of the clusters. It is therefore not clear if such
a relation is spurious (due to the small number of clusters of our sample) or
real (and masked by the presence of outliers in the MvM05 sample).

\section*{Acknowledgements}

Based on observations made with the Isaac Newton Telescope operated on the island of La Palma by the Isaac Newton Group in the Spanish Observatorio del Roque de los Muchachos of the Instituto de Astrof\'isica de Canarias and with 2.2\,m ESO telescope at the La Silla Observatory under programme IDs 082.B-0386, 084.B-0666 and 085.B-0765. We warmly thank the anonymous referee for his/her helpful comments and suggestions. We also thank Liliya L. R. Williams for her comments. MG acknowledges financial support from the Royal Society. JP acknowledges support from the STFC-funded Galaxy Formation and Evolution programme at the IoA.

\def\jnl@style{\it}                       
\def\mnref@jnl#1{{\jnl@style#1}}          

\def\aj{\mnref@jnl{AJ}}                   
\def\apj{\mnref@jnl{ApJ}}                 
\def\apjl{\mnref@jnl{ApJL}}               
\def\aap{\mnref@jnl{A\&A}}                
\def\mnras{\mnref@jnl{MNRAS}}             
\def\nat{\mnref@jnl{Nat.}}                
\def\iaucirc{\mnref@jnl{IAU~Circ.}}       
\def\atel{\mnref@jnl{ATel}}               
\def\iausymp{\mnref@jnl{IAU~Symp.}}       

\def\pasp{\mnref@jnl{PASP}}                             
\def\araa{\mnref@jnl{ARA\&A}}                           
\def\apjs{\mnref@jnl{ApJS}}                             
\def\aapr{\mnref@jnl{A\&A Rev.}}                        
\def\memsai{\mnref@jnl{\rm Mem. Soc. Astron. Italiana}} 

\nocite{1987degc.book.....S}

\bibliographystyle{mn2e}
\bibliography{biblio}

\begin{thebibliography}{}

\bibitem[\protect\citeauthoryear{{Allen}, {Moreno} \& {Pichardo}}{{Allen}
  et~al.}{2006}]{2006ApJ...652.1150A}
{Allen} C.,  {Moreno} E.,    {Pichardo} B.,  2006, \apj, 652, 1150

\bibitem[\protect\citeauthoryear{{Baraffe}, {Chabrier}, {Allard} \&
  {Hauschildt}}{{Baraffe} et~al.}{1998}]{1998A&A...337..403B}
{Baraffe} I.,  {Chabrier} G.,  {Allard} F.,    {Hauschildt} P.~H.,  1998, \aap,
  337, 403

\bibitem[\protect\citeauthoryear{{Bastian}, {Gieles}, {Goodwin}, {Trancho},
  {Smith}, {Konstantopoulos} \& {Efremov}}{{Bastian}
  et~al.}{2008}]{2008MNRAS.389..223B}
{Bastian} N.,  {Gieles} M.,  {Goodwin} S.~P.,  {Trancho} G.,  {Smith} L.~J.,
  {Konstantopoulos} I.,    {Efremov} Y.,  2008, \mnras, 389, 223

\bibitem[\protect\citeauthoryear{{Baumgardt}}{{Baumgardt}}{2001}]{2001MNRAS.325.1323B}
{Baumgardt} H.,  2001, \mnras, 325, 1323

\bibitem[\protect\citeauthoryear{{Baumgardt} \& {Makino}}{{Baumgardt} \&
  {Makino}}{2003}]{2003MNRAS.340..227B}
{Baumgardt} H.,  {Makino} J.,  2003, \mnras, 340, 227

\bibitem[\protect\citeauthoryear{{Baumgardt}, {Parmentier}, {Gieles} \&
  {Vesperini}}{{Baumgardt} et~al.}{2010}]{2010MNRAS.401.1832B}
{Baumgardt} H.,  {Parmentier} G.,  {Gieles} M.,    {Vesperini} E.,  2010,
  \mnras, 401, 1832

\bibitem[\protect\citeauthoryear{{Bellazzini}, {Ferraro} \&
  {Ibata}}{{Bellazzini} et~al.}{2003}]{2003AJ....125..188B}
{Bellazzini} M.,  {Ferraro} F.~R.,    {Ibata} R.,  2003, \aj, 125, 188

\bibitem[\protect\citeauthoryear{{Bellazzini}, {Fusi Pecci}, {Messineo},
  {Monaco} \& {Rood}}{{Bellazzini} et~al.}{2002}]{2002AJ....123.1509B}
{Bellazzini} M.,  {Fusi Pecci} F.,  {Messineo} M.,  {Monaco} L.,    {Rood}
  R.~T.,  2002, \aj, 123, 1509

\bibitem[\protect\citeauthoryear{{Bellazzini}, {Ibata} \&
  {Ferraro}}{{Bellazzini} et~al.}{2004}]{2004ASPC..327..220B}
{Bellazzini} M.,  {Ibata} R.,    {Ferraro} F.~R.,  2004, in {F.~Prada,
  D.~Martinez Delgado, \& T.~J.~Mahoney} ed., Satellites and Tidal Streams
  Vol.~327 of Astronomical Society of the Pacific Conference Series, {Globular
  Clusters in the Sgr Stream and Other Structures}.
pp 220--+

\bibitem[\protect\citeauthoryear{{Belokurov}, {Evans}, {Irwin}, {Hewett} \&
  {Wilkinson}}{{Belokurov} et~al.}{2006}]{2006ApJ...637L..29B}
{Belokurov} V.,  {Evans} N.~W.,  {Irwin} M.~J.,  {Hewett} P.~C.,    {Wilkinson}
  M.~I.,  2006, \apjl, 637, L29

\bibitem[\protect\citeauthoryear{{Binney} \& {Merrifield}}{{Binney} \&
  {Merrifield}}{1998}]{1998gaas.book.....B}
{Binney} J.,  {Merrifield} M.,  1998, {Galactic Astronomy}

\bibitem[\protect\citeauthoryear{{Borissova}, {Catelan}, {Ferraro}, {Spassova},
  {Buonanno}, {Iannicola}, {Richtler} \& {Sweigart}}{{Borissova}
  et~al.}{1999}]{Borissova1999}
{Borissova} J.,  {Catelan} M.,  {Ferraro} F.~R.,  {Spassova} N.,  {Buonanno}
  R.,  {Iannicola} G.,  {Richtler} T.,    {Sweigart} A.~V.,  1999, \aap, 343,
  813

\bibitem[\protect\citeauthoryear{{Brodie} \& {Strader}}{{Brodie} \&
  {Strader}}{2006}]{2006ARA&A..44..193B}
{Brodie} J.~P.,  {Strader} J.,  2006, \araa, 44, 193

\bibitem[\protect\citeauthoryear{{Carraro}}{{Carraro}}{2009}]{2009AJ....137.3809C}
{Carraro} G.,  2009, \aj, 137, 3809

\bibitem[\protect\citeauthoryear{{Carraro}, {Zinn} \& {Moni Bidin}}{{Carraro}
  et~al.}{2007}]{2007A&A...466..181C}
{Carraro} G.,  {Zinn} R.,    {Moni Bidin} C.,  2007, \aap, 466, 181

\bibitem[\protect\citeauthoryear{{Chernoff} \& {Shapiro}}{{Chernoff} \&
  {Shapiro}}{1987}]{1987ApJ...322..113C}
{Chernoff} D.~F.,  {Shapiro} S.~L.,  1987, \apj, 322, 113

\bibitem[\protect\citeauthoryear{{Correnti}, {Bellazzini}, {Dalessandro},
  {Mucciarelli}, {Monaco} \& {Catelan}}{{Correnti}
  et~al.}{2011}]{2011arXiv1105.2001C}
{Correnti} M.,  {Bellazzini} M.,  {Dalessandro} E.,  {Mucciarelli} A.,
  {Monaco} L.,    {Catelan} M.,  2011, arXiv:1105.2001

\bibitem[\protect\citeauthoryear{{C{\^o}t{\'e}}, {Djorgovski}, {Meylan},
  {Castro} \& {McCarthy}}{{C{\^o}t{\'e}} et~al.}{2002}]{2002ApJ...574..783C}
{C{\^o}t{\'e}} P.,  {Djorgovski} S.~G.,  {Meylan} G.,  {Castro} S.,
  {McCarthy} J.~K.,  2002, \apj, 574, 783

\bibitem[\protect\citeauthoryear{{Dehnen}, {Odenkirchen}, {Grebel} \&
  {Rix}}{{Dehnen} et~al.}{2004}]{2004AJ....127.2753D}
{Dehnen} W.,  {Odenkirchen} M.,  {Grebel} E.~K.,    {Rix} H.-W.,  2004, \aj,
  127, 2753

\bibitem[\protect\citeauthoryear{{Dinescu}, {Girard} \& {van Altena}}{{Dinescu}
  et~al.}{1999}]{1999AJ....117.1792D}
{Dinescu} D.~I.,  {Girard} T.~M.,    {van Altena} W.~F.,  1999, \aj, 117, 1792

\bibitem[\protect\citeauthoryear{{Djorgovski} \& {King}}{{Djorgovski} \&
  {King}}{1986}]{1986ApJ...305L..61D}
{Djorgovski} S.,  {King} I.~R.,  1986, \apjl, 305, L61

\bibitem[\protect\citeauthoryear{{Elson}, {Fall} \& {Freeman}}{{Elson}
  et~al.}{1987}]{1987ApJ...323...54E}
{Elson} R.~A.~W.,  {Fall} S.~M.,    {Freeman} K.~C.,  1987, \apj, 323, 54

\bibitem[\protect\citeauthoryear{{Elson}, {Freeman} \& {Lauer}}{{Elson}
  et~al.}{1989}]{1989ApJ...347L..69E}
{Elson} R.~A.~W.,  {Freeman} K.~C.,    {Lauer} T.~R.,  1989, \apjl, 347, L69

\bibitem[\protect\citeauthoryear{{Forbes} \& {Bridges}}{{Forbes} \&
  {Bridges}}{2010}]{2010MNRAS.404.1203F}
{Forbes} D.~A.,  {Bridges} T.,  2010, \mnras, 404, 1203

\bibitem[\protect\citeauthoryear{{Fregeau}, {Ivanova} \& {Rasio}}{{Fregeau}
  et~al.}{2009}]{2009ApJ...707.1533F}
{Fregeau} J.~M.,  {Ivanova} N.,    {Rasio} F.~A.,  2009, \apj, 707, 1533

\bibitem[\protect\citeauthoryear{{Fregeau} \& {Rasio}}{{Fregeau} \&
  {Rasio}}{2007}]{2007ApJ...658.1047F}
{Fregeau} J.~M.,  {Rasio} F.~A.,  2007, \apj, 658, 1047

\bibitem[\protect\citeauthoryear{{Fukushige} \& {Heggie}}{{Fukushige} \&
  {Heggie}}{2000}]{2000MNRAS.318..753F}
{Fukushige} T.,  {Heggie} D.~C.,  2000, \mnras, 318, 753

\bibitem[\protect\citeauthoryear{{Gao}, {Goodman}, {Cohn} \& {Murphy}}{{Gao}
  et~al.}{1991}]{1991ApJ...370..567G}
{Gao} B.,  {Goodman} J.,  {Cohn} H.,    {Murphy} B.,  1991, \apj, 370, 567

\bibitem[\protect\citeauthoryear{{Gieles} \& {Baumgardt}}{{Gieles} \&
  {Baumgardt}}{2008}]{2008MNRAS.389L..28G}
{Gieles} M.,  {Baumgardt} H.,  2008, \mnras, 389, L28

\bibitem[\protect\citeauthoryear{{Gieles}, {Baumgardt}, {Heggie} \&
  {Lamers}}{{Gieles} et~al.}{2010}]{2010MNRAS.408L..16G}
{Gieles} M.,  {Baumgardt} H.,  {Heggie} D.~C.,    {Lamers} H.~J.~G.~L.~M.,
  2010, \mnras, 408, L16

\bibitem[\protect\citeauthoryear{{Gieles}, {Heggie} \& {Zhao}}{{Gieles}
  et~al.}{2011}]{G11}
{Gieles} M.,  {Heggie} D.~C.,    {Zhao} H.,  2011, \mnras, 413, 2509

\bibitem[\protect\citeauthoryear{{Giersz} \& {Heggie}}{{Giersz} \&
  {Heggie}}{1994}]{1994MNRAS.268..257G}
{Giersz} M.,  {Heggie} D.~C.,  1994, \mnras, 268, 257

\bibitem[\protect\citeauthoryear{{Giersz} \& {Spurzem}}{{Giersz} \&
  {Spurzem}}{1994}]{1994MNRAS.269..241G}
{Giersz} M.,  {Spurzem} R.,  1994, \mnras, 269, 241

\bibitem[\protect\citeauthoryear{{Gnedin} \& {Ostriker}}{{Gnedin} \&
  {Ostriker}}{1997}]{1997ApJ...474..223G}
{Gnedin} O.~Y.,  {Ostriker} J.~P.,  1997, \apj, 474, 223

\bibitem[\protect\citeauthoryear{{Gnedin} \& {Ostriker}}{{Gnedin} \&
  {Ostriker}}{1999}]{1999ApJ...513..626G}
{Gnedin} O.~Y.,  {Ostriker} J.~P.,  1999, \apj, 513, 626

\bibitem[\protect\citeauthoryear{{Goodman}}{{Goodman}}{1984}]{1984ApJ...280..298G}
{Goodman} J.,  1984, \apj, 280, 298

\bibitem[\protect\citeauthoryear{{Harris}}{{Harris}}{1996}]{1996AJ....112.1487H}
{Harris} W.~E.,  1996, \aj, 112, 1487

\bibitem[\protect\citeauthoryear{{Harris}}{{Harris}}{2010}]{2010arXiv1012.3224H}
{Harris} W.~E.,  2010, arXiv:1012.3224

\bibitem[\protect\citeauthoryear{{H{\'e}non}}{{H{\'e}non}}{1961}]{1961AnAp...24..369H}
{H{\'e}non} M.,  1961, Annales d'Astrophysique, 24, 369

\bibitem[\protect\citeauthoryear{{H{\'e}non}}{{H{\'e}non}}{1965}]{1965AnAp...28...62H}
{H{\'e}non} M.,  1965, Annales d'Astrophysique, 28, 62

\bibitem[\protect\citeauthoryear{{Hut}, {McMillan}, {Goodman}, {Mateo},
  {Phinney}, {Pryor}, {Richer}, {Verbunt} \& {Weinberg}}{{Hut}
  et~al.}{1992}]{1992PASP..104..981H}
{Hut} P.,  {McMillan} S.,  {Goodman} J.,  {Mateo} M.,  {Phinney} E.~S.,
  {Pryor} C.,  {Richer} H.~B.,  {Verbunt} F.,    {Weinberg} M.,  1992, \pasp,
  104, 981

\bibitem[\protect\citeauthoryear{{Innanen}, {Harris} \& {Webbink}}{{Innanen}
  et~al.}{1983}]{1983AJ.....88..338I}
{Innanen} K.~A.,  {Harris} W.~E.,    {Webbink} R.~F.,  1983, \aj, 88, 338

\bibitem[\protect\citeauthoryear{{Johnston}, {Sigurdsson} \&
  {Hernquist}}{{Johnston} et~al.}{1999}]{1999MNRAS.302..771J}
{Johnston} K.~V.,  {Sigurdsson} S.,    {Hernquist} L.,  1999, \mnras, 302, 771

\bibitem[\protect\citeauthoryear{{Jordi} \& {Grebel}}{{Jordi} \&
  {Grebel}}{2010}]{2010A&A...522A..71J}
{Jordi} K.,  {Grebel} E.~K.,  2010, \aap, 522, A71+

\bibitem[\protect\citeauthoryear{{King}}{{King}}{1962}]{1962AJ.....67..471K}
{King} I.,  1962, \aj, 67, 471

\bibitem[\protect\citeauthoryear{{King}}{{King}}{1966}]{1966AJ.....71...64K}
{King} I.~R.,  1966, \aj, 71, 64

\bibitem[\protect\citeauthoryear{{K{\"u}pper}, {Kroupa}, {Baumgardt} \&
  {Heggie}}{{K{\"u}pper} et~al.}{2010}]{2010MNRAS.407.2241K}
{K{\"u}pper} A.~H.~W.,  {Kroupa} P.,  {Baumgardt} H.,    {Heggie} D.~C.,  2010,
  \mnras, 407, 2241

\bibitem[\protect\citeauthoryear{{Landolt}}{{Landolt}}{1992}]{1992AJ....104..340L}
{Landolt} A.~U.,  1992, \aj, 104, 340

\bibitem[\protect\citeauthoryear{{Larsen}}{{Larsen}}{2004}]{2004A&A...416..537L}
{Larsen} S.~S.,  2004, \aap, 416, 537

\bibitem[\protect\citeauthoryear{{Lauchner}, {Powell} Jr. \&
  {Wilhelm}}{{Lauchner} et~al.}{2006}]{2006ApJ...651L..33L}
{Lauchner} A.,  {Powell} Jr. W.~L.,    {Wilhelm} R.,  2006, \apjl, 651, L33

\bibitem[\protect\citeauthoryear{{Lee} \& {Ostriker}}{{Lee} \&
  {Ostriker}}{1987}]{1987ApJ...322..123L}
{Lee} H.~M.,  {Ostriker} J.~P.,  1987, \apj, 322, 123

\bibitem[\protect\citeauthoryear{{Lee}, {Lee} \& {Sung}}{{Lee}
  et~al.}{2006}]{2006MNRAS.367..646L}
{Lee} K.~H.,  {Lee} H.~M.,    {Sung} H.,  2006, \mnras, 367, 646

\bibitem[\protect\citeauthoryear{{Leon}, {Meylan} \& {Combes}}{{Leon}
  et~al.}{2000}]{Leon2000}
{Leon} S.,  {Meylan} G.,    {Combes} F.,  2000, \aap, 359, 907

\bibitem[\protect\citeauthoryear{{Mackey} \& {Gilmore}}{{Mackey} \&
  {Gilmore}}{2003a}]{2003MNRAS.338..120M}
{Mackey} A.~D.,  {Gilmore} G.~F.,  2003a, \mnras, 338, 120

\bibitem[\protect\citeauthoryear{{Mackey} \& {Gilmore}}{{Mackey} \&
  {Gilmore}}{2003b}]{2003MNRAS.338...85M}
{Mackey} A.~D.,  {Gilmore} G.~F.,  2003b, \mnras, 338, 85

\bibitem[\protect\citeauthoryear{{Mackey}, {Wilkinson}, {Davies} \&
  {Gilmore}}{{Mackey} et~al.}{2008}]{2008MNRAS.386...65M}
{Mackey} A.~D.,  {Wilkinson} M.~I.,  {Davies} M.~B.,    {Gilmore} G.~F.,  2008,
  \mnras, 386, 65

\bibitem[\protect\citeauthoryear{{Mackey et al.}}{{Mackey et
  al.}}{2010}]{2010MNRAS.401..533M}
{Mackey et al.} 2010, \mnras, 401, 533

\bibitem[\protect\citeauthoryear{{Ma{\'{\i}}z-Apell{\'a}niz}}{{Ma{\'{\i}}z-Apell{\'a}niz}}{2001}]{2001ApJ...563..151M}
{Ma{\'{\i}}z-Apell{\'a}niz} J.,  2001, \apj, 563, 151

\bibitem[\protect\citeauthoryear{{Marigo}, {Girardi}, {Bressan}, {Groenewegen},
  {Silva} \& {Granato}}{{Marigo} et~al.}{2008}]{2008A&A...482..883M}
{Marigo} P.,  {Girardi} L.,  {Bressan} A.,  {Groenewegen} M.~A.~T.,  {Silva}
  L.,    {Granato} G.~L.,  2008, \aap, 482, 883

\bibitem[\protect\citeauthoryear{{Mar{\'{\i}}n-Franch}, {Aparicio}, {Piotto},
  {Rosenberg}, {Chaboyer}, {Sarajedini}, {Siegel}, {Anderson}, {Bedin},
  {Dotter}, {Hempel}, {King}, {Majewski}, {Milone}, {Paust} \&
  {Reid}}{{Mar{\'{\i}}n-Franch} et~al.}{2009}]{2009ApJ...694.1498M}
{Mar{\'{\i}}n-Franch} A.,  {Aparicio} A.,  {Piotto} G.,  {Rosenberg} A.,
  {Chaboyer} B.,  {Sarajedini} A.,  {Siegel} M.,  {Anderson} J.,  {Bedin}
  L.~R.,  {Dotter} A.,  {Hempel} M.,  {King} I.,  {Majewski} S.,  {Milone}
  A.~P.,  {Paust} N.,    {Reid} I.~N.,  2009, \apj, 694, 1498

\bibitem[\protect\citeauthoryear{{Martin}, {Ibata}, {Bellazzini}, {Irwin},
  {Lewis} \& {Dehnen}}{{Martin} et~al.}{2004}]{2004MNRAS.348...12M}
{Martin} N.~F.,  {Ibata} R.~A.,  {Bellazzini} M.,  {Irwin} M.~J.,  {Lewis}
  G.~F.,    {Dehnen} W.,  2004, \mnras, 348, 12

\bibitem[\protect\citeauthoryear{{Mart{\'{\i}}nez-Delgado}, {Dinescu}, {Zinn},
  {Tutsoff}, {C{\^o}t{\'e}} \& {Boyarchuck}}{{Mart{\'{\i}}nez-Delgado}
  et~al.}{2004}]{2004ASPC..327..255M}
{Mart{\'{\i}}nez-Delgado} D.,  {Dinescu} D.~I.,  {Zinn} R.,  {Tutsoff} A.,
  {C{\^o}t{\'e}} P.,    {Boyarchuck} A.,  2004, in {F.~Prada, D.~Martinez
  Delgado, \& T.~J.~Mahoney} ed., Satellites and Tidal Streams Vol.~327 of
  Astronomical Society of the Pacific Conference Series, {Mapping Tidal Streams
  around Galactic Globular Clusters}.
pp 255--+

\bibitem[\protect\citeauthoryear{{McLaughlin}}{{McLaughlin}}{2000}]{2000ApJ...539..618M}
{McLaughlin} D.~E.,  2000, \apj, 539, 618

\bibitem[\protect\citeauthoryear{{McLaughlin} \& {van der Marel}}{{McLaughlin}
  \& {van der Marel}}{2005}]{2005ApJS..161..304M}
{McLaughlin} D.~E.,  {van der Marel} R.~P.,  2005, \apjs, 161, 304

\bibitem[\protect\citeauthoryear{{Meylan} \& {Heggie}}{{Meylan} \&
  {Heggie}}{1997}]{1997A&ARv...8....1M}
{Meylan} G.,  {Heggie} D.~C.,  1997, \aapr, 8, 1

\bibitem[\protect\citeauthoryear{{Michie}}{{Michie}}{1963}]{1963MNRAS.126..499M}
{Michie} R.~W.,  1963, \mnras, 126, 499

\bibitem[\protect\citeauthoryear{{Milone}, {Piotto}, {Bedin} \&
  {Sarajedini}}{{Milone} et~al.}{2008}]{2008MmSAI..79..623M}
{Milone} A.~P.,  {Piotto} G.,  {Bedin} L.~R.,    {Sarajedini} A.,  2008,
  \memsai, 79, 623

\bibitem[\protect\citeauthoryear{{Niederste-Ostholt}, {Belokurov}, {Evans},
  {Koposov}, {Gieles} \& {Irwin}}{{Niederste-Ostholt}
  et~al.}{2010}]{2010MNRAS.408L..66N}
{Niederste-Ostholt} M.,  {Belokurov} V.,  {Evans} N.~W.,  {Koposov} S.,
  {Gieles} M.,    {Irwin} M.~J.,  2010, \mnras, 408, L66

\bibitem[\protect\citeauthoryear{{Noyola} \& {Gebhardt}}{{Noyola} \&
  {Gebhardt}}{2006}]{2006AJ....132..447N}
{Noyola} E.,  {Gebhardt} K.,  2006, \aj, 132, 447

\bibitem[\protect\citeauthoryear{{Odenkirchen}, {Grebel}, {Dehnen}, {Rix},
  {Yanny}, {Newberg}, {Rockosi}, {Mart{\'{\i}}nez-Delgado}, {Brinkmann} \&
  {Pier}}{{Odenkirchen} et~al.}{2003}]{2003AJ....126.2385O}
{Odenkirchen} M.,  {Grebel} E.~K.,  {Dehnen} W.,  {Rix} H.,  {Yanny} B.,
  {Newberg} H.~J.,  {Rockosi} C.~M.,  {Mart{\'{\i}}nez-Delgado} D.,
  {Brinkmann} J.,    {Pier} J.~R.,  2003, \aj, 126, 2385

\bibitem[\protect\citeauthoryear{{Odenkirchen et al.}}{{Odenkirchen et
  al.}}{2001}]{2001ApJ...548L.165O}
{Odenkirchen et al.} 2001, \apjl, 548, L165

\bibitem[\protect\citeauthoryear{{Oh}, {Lin} \& {Aarseth}}{{Oh}
  et~al.}{1995}]{1995ApJ...442..142O}
{Oh} K.~S.,  {Lin} D.~N.~C.,    {Aarseth} S.~J.,  1995, \apj, 442, 142

\bibitem[\protect\citeauthoryear{{Olszewski}, {Saha}, {Knezek}, {Subramaniam},
  {de Boer} \& {Seitzer}}{{Olszewski} et~al.}{2009}]{2009AJ....138.1570O}
{Olszewski} E.~W.,  {Saha} A.,  {Knezek} P.,  {Subramaniam} A.,  {de Boer} T.,
    {Seitzer} P.,  2009, \aj, 138, 1570

\bibitem[\protect\citeauthoryear{{Ostriker}, {Spitzer} \&
  {Chevalier}}{{Ostriker} et~al.}{1972}]{1972ApJ...176L..51O}
{Ostriker} J.~P.,  {Spitzer} L.~J.,    {Chevalier} R.~A.,  1972, \apjl, 176,
  L51+

\bibitem[\protect\citeauthoryear{{Pe{\~n}arrubia}, {Navarro}, {McConnachie} \&
  {Martin}}{{Pe{\~n}arrubia} et~al.}{2009}]{2009ApJ...698..222P}
{Pe{\~n}arrubia} J.,  {Navarro} J.~F.,  {McConnachie} A.~W.,    {Martin} N.~F.,
   2009, \apj, 698, 222

\bibitem[\protect\citeauthoryear{{Plummer}}{{Plummer}}{1911}]{1911MNRAS..71..460P}
{Plummer} H.~C.,  1911, \mnras, 71, 460

\bibitem[\protect\citeauthoryear{{Searle} \& {Zinn}}{{Searle} \&
  {Zinn}}{1978}]{1978ApJ...225..357S}
{Searle} L.,  {Zinn} R.,  1978, \apj, 225, 357

\bibitem[\protect\citeauthoryear{{Sollima}}{{Sollima}}{2008}]{2008MNRAS.388..307S}
{Sollima} A.,  2008, \mnras, 388, 307

\bibitem[\protect\citeauthoryear{{Sollima}, {Beccari}, {Ferraro}, {Fusi Pecci}
  \& {Sarajedini}}{{Sollima} et~al.}{2007}]{2007MNRAS.380..781S}
{Sollima} A.,  {Beccari} G.,  {Ferraro} F.~R.,  {Fusi Pecci} F.,
  {Sarajedini} A.,  2007, \mnras, 380, 781

\bibitem[\protect\citeauthoryear{{Sollima}, {Carballo-Bello}, {Beccari},
  {Ferraro}, {Pecci} \& {Lanzoni}}{{Sollima}
  et~al.}{2010}]{2010MNRAS.401..577S}
{Sollima} A.,  {Carballo-Bello} J.~A.,  {Beccari} G.,  {Ferraro} F.~R.,
  {Pecci} F.~F.,    {Lanzoni} B.,  2010, \mnras, 401, 577

\bibitem[\protect\citeauthoryear{{Sollima}, {Mart{\'{\i}}nez-Delgado},
  {Valls-Gabaud} \& {Pe{\~n}arrubia}}{{Sollima}
  et~al.}{2011}]{2011ApJ...726...47S}
{Sollima} A.,  {Mart{\'{\i}}nez-Delgado} D.,  {Valls-Gabaud} D.,
  {Pe{\~n}arrubia} J.,  2011, \apj, 726, 47

\bibitem[\protect\citeauthoryear{{Spitzer}}{{Spitzer}}{1987}]{1987degc.book.....S}
{Spitzer} L.,  1987, {Dynamical evolution of globular clusters}.
Princeton, NJ, Princeton University Press, 1987, 191 p.

\bibitem[\protect\citeauthoryear{{Spitzer} \& {Hart}}{{Spitzer} \&
  {Hart}}{1971}]{1971ApJ...164..399S}
{Spitzer} L.~J.,  {Hart} M.~H.,  1971, \apj, 164, 399

\bibitem[\protect\citeauthoryear{{Stetson}}{{Stetson}}{1987}]{1987PASP...99..191S}
{Stetson} P.~B.,  1987, \pasp, 99, 191

\bibitem[\protect\citeauthoryear{{Testa}, {Zaggia}, {Andreon}, {Longo},
  {Scaramella}, {Djorgovski} \& {de Carvalho}}{{Testa}
  et~al.}{2000}]{2000A&A...356..127T}
{Testa} V.,  {Zaggia} S.~R.,  {Andreon} S.,  {Longo} G.,  {Scaramella} R.,
  {Djorgovski} S.~G.,    {de Carvalho} R.,  2000, \aap, 356, 127

\bibitem[\protect\citeauthoryear{{Trager}, {King} \& {Djorgovski}}{{Trager}
  et~al.}{1995}]{1995AJ....109..218T}
{Trager} S.~C.,  {King} I.~R.,    {Djorgovski} S.,  1995, \aj, 109, 218

\bibitem[\protect\citeauthoryear{{Vesperini}}{{Vesperini}}{1998}]{1998MNRAS.299.1019V}
{Vesperini} E.,  1998, \mnras, 299, 1019

\bibitem[\protect\citeauthoryear{{Vesperini} \& {Heggie}}{{Vesperini} \&
  {Heggie}}{1997}]{1997MNRAS.289..898V}
{Vesperini} E.,  {Heggie} D.~C.,  1997, \mnras, 289, 898

\bibitem[\protect\citeauthoryear{{Walker et al.}}{{Walker et
  al.}}{2011}]{2011arXiv1103.4144W}
{Walker et al.} 2011, \mnras, 415, 643

\bibitem[\protect\citeauthoryear{{Wilkinson}, {Hurley}, {Mackey}, {Gilmore} \&
  {Tout}}{{Wilkinson} et~al.}{2003}]{2003MNRAS.343.1025W}
{Wilkinson} M.~I.,  {Hurley} J.~R.,  {Mackey} A.~D.,  {Gilmore} G.~F.,
  {Tout} C.~A.,  2003, \mnras, 343, 1025

\bibitem[\protect\citeauthoryear{{Wilson}}{{Wilson}}{1975}]{1975AJ.....80..175W}
{Wilson} C.~P.,  1975, \aj, 80, 175

\bibitem[\protect\citeauthoryear{{Woolley} \& {Robertson}}{{Woolley} \&
  {Robertson}}{1956}]{1956MNRAS.116..288W}
{Woolley} R.~V.~D.~R.,  {Robertson} D.~A.,  1956, \mnras, 116, 288

\bibitem[\protect\citeauthoryear{{Zepf} \& {Ashman}}{{Zepf} \&
  {Ashman}}{1993}]{1993MNRAS.264..611Z}
{Zepf} S.~E.,  {Ashman} K.~M.,  1993, \mnras, 264, 611

\bibitem[\protect\citeauthoryear{{Zhao}}{{Zhao}}{1996}]{1996MNRAS.278..488Z}
{Zhao} H.,  1996, \mnras, 278, 488

\bibitem[\protect\citeauthoryear{{Zinn}}{{Zinn}}{1993}]{1993ASPC...48...38Z}
{Zinn} R.,  1993, in {G.~H.~Smith \& J.~P.~Brodie} ed., The Globular
  Cluster-Galaxy Connection Vol.~48 of Astronomical Society of the Pacific
  Conference Series, {The Galactic Halo Cluster Systems: Evidence for
  Accretion}.
pp 38--+

\end{thebibliography}

\clearpage

\label{lastpage}

\end{document}